\documentclass[pdflatex,sn-nature]{sn-jnl}
\usepackage{amsmath,amssymb}
\usepackage{booktabs,graphicx,array}
\graphicspath{{figures/}}
\expandafter\def\expandafter\abstractfont\expandafter{\abstractfont\unboldmath}

\begin{document}
\title[Reusing formal Vlasov--Maxwell proofs]{Reusing formal Vlasov--Maxwell proofs to audit plasma simulations}
\author[1]{\fnm{Bowen} \sur{Zhu}}
\author*[1]{\fnm{Jian} \sur{Wu}}\email{jxjawj@mail.xjtu.edu.cn}
\affil[1]{\orgname{State Key Laboratory of Electrical Insulation and Power Equipment, Xi'an Jiaotong University},
\orgaddress{\city{Xi'an}, \postcode{710049}, \country{China}}}
\abstract{Three-dimensional relativistic plasma equations rarely admit exact nonlinear solutions. Code verification therefore relies on global energy and mass conservation. Here we turn machine-checked lemmas from a Lean~4 Vlasov--Maxwell library into initial-data oracles that audit production plasma codes. Outside a compact plasma, finite propagation speed fixes exterior fields and box-energy flux, and a discrete Yee-grid identity forbids spurious axial magnetic fields from divergence-free initial data. In WarpX and Smilei, these checks caught domain-truncation and grid-initialization errors missed by energy conservation and uncovered an upstream bug. Inside a collisionless plasma, Liouville's theorem bounds phase-space mass concentration, flagging unphysical clumping from energy-conserving particle merging and compressive remapping. Arbitrary-precision enclosures evaluate these references in strongly absorbing pulse-driven and pulse-free plasma regimes. Together, AI coding agents and Lean~4 turn classical theorems into practical verification tools, letting physicists audit minimal statements and assumptions rather than full proofs.}
\maketitle

\section*{Introduction}

Once particles and electromagnetic fields evolve nonlinearly in three dimensions (3D), the relativistic Vlasov--Maxwell equations rarely admit closed-form solutions~\cite{glasseystrauss1986,oconnor2021}. Production plasma codes are therefore tested with global energy and charge conservation, simplified linear limits, manufactured sources, or cross-code comparisons~\cite{birdsall2004,roache2002,oberkampf2010,riva2014}. Each of these checks misses distinct classes of error: opposing local errors cancel in global sums, manufactured sources alter physical particle--field coupling and can introduce derivation mistakes into the test, and independent particle-in-cell (PIC) codes often share the same grid or resampling artifacts. More rigorous analytical benchmarks and mechanized proofs have so far been limited to specialized settings such as rigid bunches, bounded magnetized plasmas, linear time-harmonic Maxwell problems, or centered finite-difference wave codes~\cite{haletucker,gueroult,chenmaxwell,boldo2013}.

\begin{figure}[!tbp]
\centering\includegraphics[width=\linewidth]{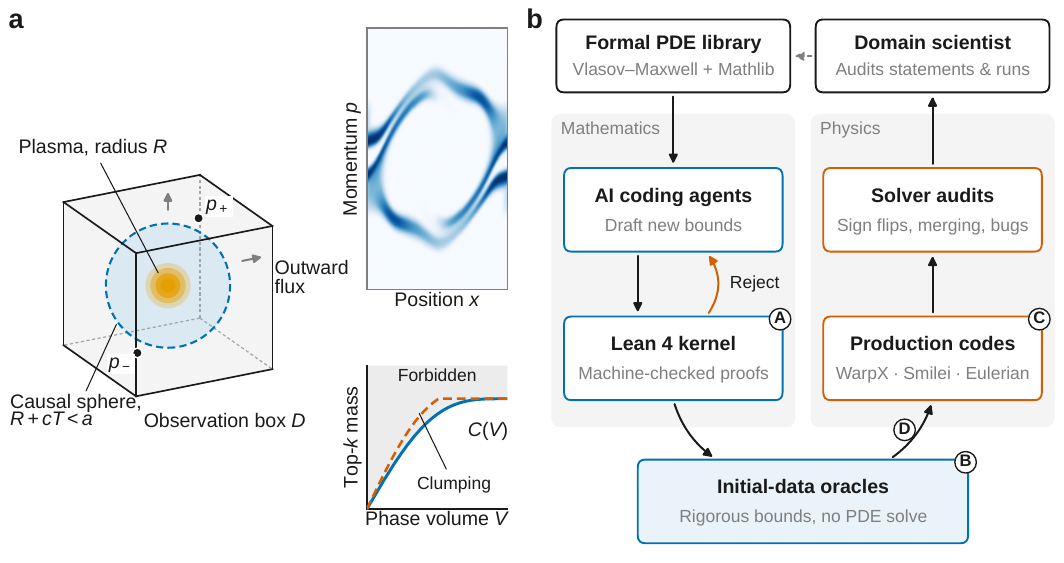}
\caption{\textbf{Initial-data bounds and workflow for auditing plasma solvers.}
\textbf{a}, Exterior and interior bounds from initial data. Left: because plasma currents remain within the expanding causal sphere ($R+cT<a$), fields at exterior probes ($p_+$, $p_-$) and outward energy flux across the observation box $D$ follow from the free-space evolution of the initial fields. Right: phase-space filamentation in $(x,p)$ (top) and cumulative mass in the densest cells versus phase volume $V$ (bottom); Liouville volume preservation keeps exact dynamics below the initial-data cap $C(V)$ (solid curve), whereas artificial clumping enters the forbidden region (dashed curve).
\textbf{b}, AI agents draft bounds that the Lean~4 kernel checks; physicists audit the theorem statements and apply the resulting oracles to production codes. Circled letters mark the evidence tier of each step: A, kernel-proved; B, rigorous numerical enclosure; C, solver measurement; D, assumed premise (smooth solutions and the continuum-to-grid bridge).}\label{fig:trust-boundary}
\end{figure}

Public machine-checked libraries in Lean~4~\cite{demoura2021,mathlib2020} have expanded rapidly through artificial-intelligence (AI)-assisted formalization~\cite{openaimath}. Recent work further shows that recombining established elementary tools can resolve open problems, from counterexamples to the Jacobian conjecture~\cite{gao2026jacobian,vandenessen2026jacobian} to reusable lemmas mined from large proof corpora~\cite{xie}. Here we draw on OpenAI's Lean~4 formalization of global classical existence for the 3D relativistic Vlasov--Maxwell system~\cite{openairvm,openaimath}. We choose this library because, to establish global existence, it assembles intermediate lemmas capturing exact physical properties of the governing partial differential equations (PDEs). We reuse two of these properties (Fig.~\ref{fig:trust-boundary}a). Outside an initially compact plasma, finite wave and particle propagation speeds fix exterior probe fields and enclosing-box energy change to the source-free Maxwell evolution of the initial fields until a light-speed signal from the plasma reaches the probes or box boundary~\cite{openairvm,bigorgne}. Inside an interacting collisionless plasma, Lorentz transport and Liouville's theorem bound the mass that can accumulate in any given phase-space volume over long runs in closed, periodic, or open domains~\cite{openairvm,tremaine1986,dehnen2005}.

Both properties are classical on their own, yet turning them into executable solver checks requires combining analytical bounds, staggered-mesh identities, and outward-rounded interval arithmetic that simulation groups seldom assemble by hand. In our workflow (Fig.~\ref{fig:trust-boundary}b), pairing the upstream Lean~4 library with AI coding agents bridges that gap: the agents derive the needed oracle inequalities, and the proof kernel verifies every step from the stated premises. Physicists then audit the theorem statements, supporting definitions, and continuum-to-discrete assumptions, organized into four evidence tiers (Fig.~\ref{fig:trust-boundary}b and Supplementary Table~S1). From those proved bounds, we construct outward-rounded interval oracles~\cite{rump2010,tucker2011,gomezserrano2019,johansson2017arb} and apply them to WarpX~\cite{warpx,vay2018,fedeli2022,chen2020}, Smilei~\cite{smilei}, and an Eulerian solver. Across exterior, interior, and strongly nonlinear 3D benchmarks, these checks catch compact-domain truncation, staggered-grid initialization errors, an upstream divergence-cleaning bug, and unphysical clumping from energy-conserving particle merging.

\section*{Results}

\subsection*{Exterior references expose setup and solver defects}

\begin{figure}[!tbp]
\centering\includegraphics[width=\linewidth]{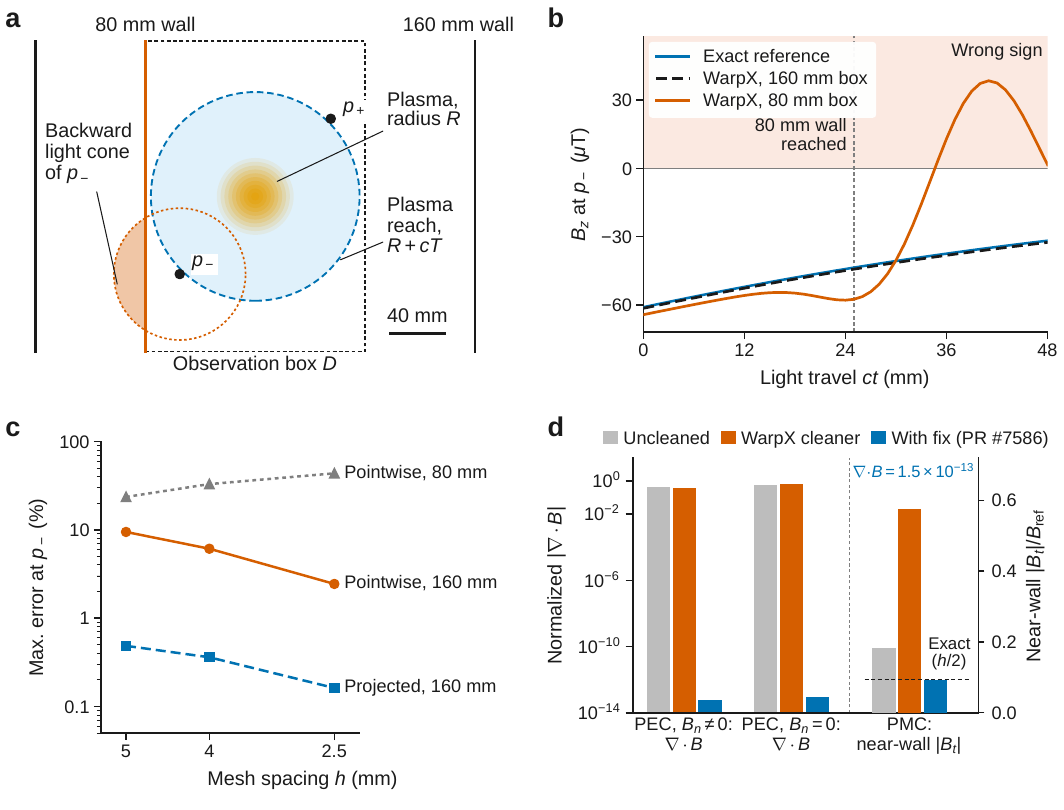}
\caption{\textbf{Exterior references identify domain truncation, staggered-grid initialization error and a divergence-cleaner bug.}
\textbf{a}, Cross-section through the plasma and probes. The backward light cone of probe $p_-$ crosses the wall of the 80~mm computational box but not that of the 160~mm box.
\textbf{b}, Magnetic field $B_z$ at $p_-$ in WarpX versus light travel $ct$. On the 80~mm box, $B_z$ reverses sign after the wall enters the probe's backward light cone (dashed line) even though box-energy balance holds within 0.5\%; the 160~mm run tracks the exact reference.
\textbf{c}, Maximum relative error at $p_-$ in Smilei versus mesh spacing $h$ for pointwise initial fields on 80~mm and 160~mm domains and divergence-free (projected) initial fields on the 160~mm domain.
\textbf{d}, Normalized residual magnetic divergence ($\nabla\cdot B$, left axis) for two perfect-electric-conductor (PEC) wall profiles ($B_n\neq0$ and $B_n=0$) and first-cell near-wall tangential magnetic field ($B_t$, right axis) for a perfect-magnetic-conductor (PMC) wall profile, comparing uncleaned fields, stock WarpX cleaning, and the fix in Pull Request \#7586; the dashed line marks the exact divergence-free near-wall value (normalization in Supplementary Note~2).}\label{fig:exterior-diag}
\end{figure}

Charged particles move slower than light, so currents from a plasma initially inside radius $R$ cannot reach beyond $R+ct$ by time $t$. Because electromagnetic waves also propagate at speed $c$, any probe or surface beyond this expanding sphere sees only the free-space Maxwell evolution of the initial fields (Fig.~\ref{fig:exterior-diag}a). Consider an observation cube $D=[-a,a]^3$ whose boundary stays outside the expanding sphere up to a horizon time $T$ ($R+cT<a$), containing total energy
\begin{equation}
 H(t)=K(t)+\frac{\epsilon_0}{2}\int_D
       \left(|E(t,x)|^2+c^2|B(t,x)|^2\right)\,\mathrm{d} x ,
 \label{eq:campaign-energy}
\end{equation}
where $K(t)$ is the particle kinetic energy inside $D$. Building on the formal Vlasov--Maxwell library~\cite{openairvm}, we prove in Lean~4 that for smooth nonlinear solutions (Tier~D), the fields on $\partial D$ and at exterior probes equal the free-space evolution $(E_v,B_v)$ of the full initial fields. Because no particles cross $\partial D$ and the boundary fields are known, Poynting's theorem fixes the energy change in $D$:
\begin{equation}
 H(t)-H(0)=-\int_0^t\int_{\partial D}
       (E_v\times B_v)\cdot\nu\,\frac{\mathrm{d} A\,\mathrm{d} s}{\mu_0}
       =U_v(t)-U_v(0),
 \label{eq:campaign-flux}
\end{equation}
where $\nu$ is the outward unit normal on $\partial D$ and $U_v$ is the field energy of $(E_v,B_v)$ inside $D$. Both the exterior field equality and equation~\eqref{eq:campaign-flux} hold even while particles and fields exchange energy nonlinearly inside $D$.

We simulated a compact positron cloud crossing a toroidal magnetic pulse in WarpX (Methods); symmetry requires probes on opposite sides of the cloud to record magnetic fields of opposite sign. On the smaller of two domains, the field at one probe reversed sign midway through every run even though box-energy balance held within 0.5\% (Fig.~\ref{fig:exterior-diag}b). Comparing the probe signal with the exterior reference showed an initial offset from domain truncation followed by a sharp sign reversal once the domain wall enters the probe's backward light cone. The reversal persisted under mesh refinement, tighter solver tolerances and a thicker absorbing layer~\cite{berenger1994}; doubling the domain restored the correct sign (Supplementary Note~2 and Supplementary Fig.~S1).

Smilei recovered the correct probe signs on the larger domain but carried a probe error near 10\% that did not shrink with more particles. Even when the magnetic pulse evolved alone in vacuum, the solver generated spurious axial magnetic lobes absent from the exact solution (Supplementary Fig.~S2). Sampling a divergence-free field point by point onto staggered Yee faces~\cite{yee} leaves a small discrete divergence, which excites axial fields under the leapfrog update~\cite{hymanshashkov,munz2000}. We proved in Lean~4 that in source-free vacuum, zero axial field persists exactly whenever the initial discrete divergence and axial electric curl vanish, and enforced this condition by projecting the initial pulse onto the discrete divergence-free subspace. This projection removed the vacuum lobes to machine precision, cut the coupled-plasma probe error from 9.45\% to 0.49\%, and preserved the initial field energy (Fig.~\ref{fig:exterior-diag}c). Initializing instead from a discrete vector potential also removed the lobes, but shifted the initial energy in a way caught by the exterior energy reference (Supplementary Note~2).

We next checked whether WarpX's built-in divergence cleaner enforces this discrete divergence condition at initialization. Measuring the divergence after cleaning showed that the cleaner barely reduced it in conducting-wall boxes and increased it for some field profiles (Fig.~\ref{fig:exterior-diag}d). We traced this to two C++ defects present in recent WarpX releases: the potential boundary conditions for perfect electric and perfect magnetic conductor (PEC and PMC) walls were swapped relative to those required for cleaning a magnetic field, and boundary ghost cells were not updated before the correction was applied (Supplementary Note~2). Our upstream fix, submitted as WarpX Pull Request \#7586~\cite{warpxpr7586}, lowers the residual divergence to round-off level in PEC and PMC tests, restores the exact near-wall tangential field on PMC boundaries, and leaves electrostatic behavior unchanged.

\subsection*{Phase-volume bounds detect remapping and merging clumping}

Exterior references break down once a light-speed signal from the plasma reaches the probes and do not constrain the interior; a second initial-data bound applies directly inside any collisionless plasma. Under relativistic Vlasov--Maxwell dynamics, each phase-space parcel retains its phase-space density along trajectories and the flow preserves phase-space volume; both properties are proved in OpenAI's library~\cite{openairvm}. Fields can stretch and fold a dense region, and crossing beams can raise the density in ordinary space, but no set of phase-space cells can ever hold more particles than the densest part of the initial plasma could fit into the same volume~\cite{tremaine1986,dehnen2005}. This gives a cap computed from the normalized initial distribution $f_0(z)$ alone,
\begin{equation}
 C(V)=\inf_{\lambda\ge 0}\left\{\lambda V+
       \int (f_0(z)-\lambda)_+\,\mathrm{d}z\right\},
 \quad M_k\le C(kv)\;\text{ for exact cell masses},
 \label{eq:occupancy}
\end{equation}
where $z=(x,p)$, $(\cdot)_+=\max(\cdot,0)$, and $M_k$ is the fraction of particles in the $k$ most populated diagnostic cells, each of phase volume $v$. Any positive excess $M_k-C(kv)$ is a lower bound on the error in the simulated particle fractions, even when the cells are chosen after inspecting the output. A simulation with $N$ markers also carries an unavoidable excess of at least $1-C(Nv)$, because $N$ point particles occupy at most $N$ cells; this sharp marker floor sets the minimum resolution baseline below which finite-marker sparsity could be mistaken for clumping. We proved 14 additional Lean~4 theorems specializing the cap and marker floor to the Maxwellian, relativistic Maxwell--J\"uttner and Gaussian distributions used in production simulations (Supplementary Note~3).

\begin{figure}[!tbp]
\centering\includegraphics[width=\linewidth]{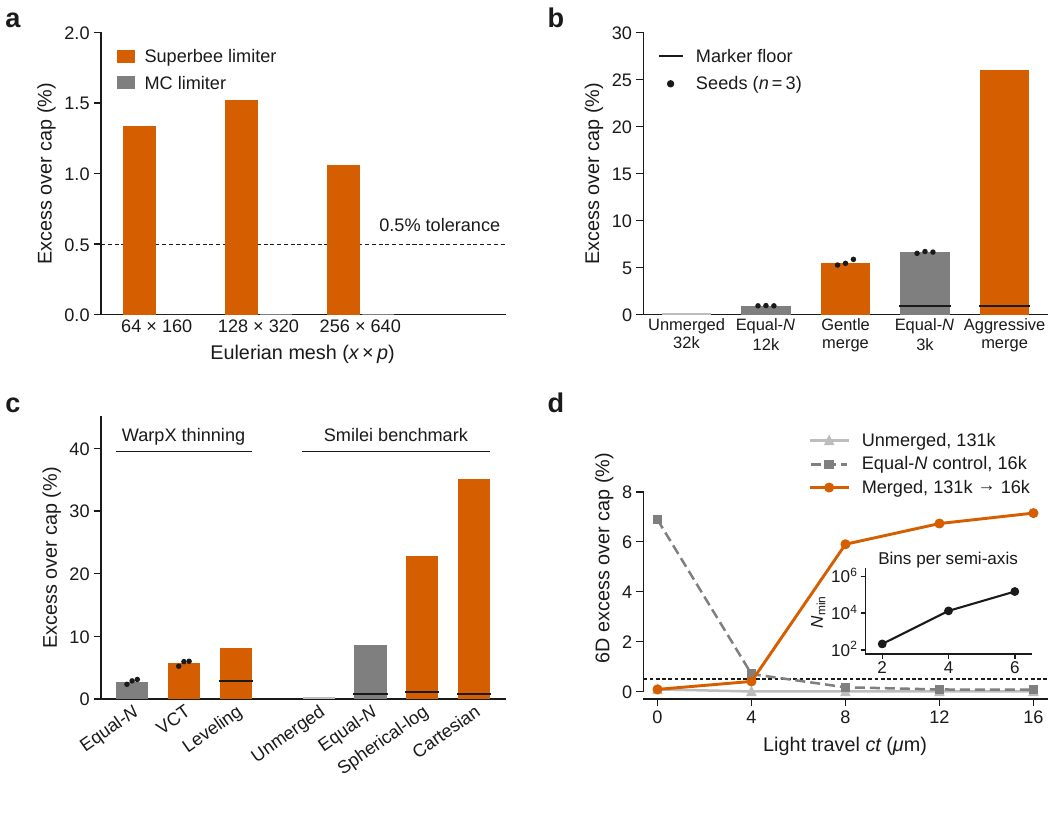}
\caption{\textbf{Phase-volume cap detects Eulerian over-compression and particle-merging clumping.}
\textbf{a}, Phase-space excess in a continuum Eulerian beam--plasma simulation across three meshes: the compressive Superbee limiter exceeds the 0.5\% tolerance (dashed line) on all three meshes, whereas the monotonized central (MC) limiter stays essentially within the cap (below 0.01\%).
\textbf{b}, Phase-space excess in a 2D Smilei Weibel-instability simulation (bars show single-seed runs for the unmerged baseline and aggressive merge, and means of $n=3$ seeds shown as dots for gentle merge and equal-$N$ controls; horizontal ticks, marker floor). Despite conserving energy and particle weight, gentle and aggressive Vranic merging produce excess far above unmerged controls matched to the same final marker count (equal-$N$).
\textbf{c}, Phase-space excess in WarpX for velocity-coincidence thinning (VCT) and equal-$N$ controls (means of $n=3$ seeds, dots; horizontal ticks, marker floor) alongside single-seed leveling thinning, and in Smilei's upstream merging benchmark for single-run unmerged, Cartesian equal-$N$, spherical-logarithmic, and Cartesian merging runs.
\textbf{d}, Six-dimensional phase-space excess versus light travel $ct$ during the 3D pulse-driven plasma run of Fig.~\ref{fig:strong-plasma}, comparing unmerged, merged ($131\mathrm{k}\to16\mathrm{k}$), and equal-$N$ control runs against the 0.5\% tolerance (dashed line); inset, minimum marker count $N_{\min}$ keeping the marker floor below 0.5\% versus diagnostic bins per semi-axis.}\label{fig:occupancy}
\end{figure}

In a continuum Eulerian simulation of a relativistic beam--plasma instability, remapping with the compressive Superbee slope limiter~\cite{sweby1984} exceeded the cap by more than 1\% on all three meshes tested, whereas the monotonized central (MC) limiter stayed essentially within it (Fig.~\ref{fig:occupancy}a). Despite this violation, Superbee conserved mass exactly, kept phase-space density positive and below its initial peak, and drifted less in energy on the coarsest mesh than MC. Because the bound is one-sided, it flags artificial compression while missing numerical diffusion, complementing global error norms (Supplementary Note~3).

We next applied the cap to particle merging, which production PIC codes use routinely to keep large simulations affordable. Existing quality checks compare a merged simulation against an expensive unmerged reference run~\cite{vranic2015,muraviev2021,gonoskov2022}. Because the cap depends only on the initial distribution, it scores a merged run directly; in our benchmarks, unmerged controls matched to the final marker count separate algorithmic clumping from finite-marker sampling noise. In a Smilei simulation of a relativistic Weibel instability, Vranic merging produced phase-space excess several times above the equal-marker controls and growing with merging aggressiveness, even though particle weight, symmetry and energy were conserved as well as in the unmerged run, and the most strongly flagged run had a 35\% error in the magnetic field (Fig.~\ref{fig:occupancy}b).

Similar excess appeared in other particle-reduction schemes (Fig.~\ref{fig:occupancy}c). In WarpX, leveling thinning failed the energy check, whereas energy-conserving velocity-coincidence thinning passed it yet still produced larger phase-space excess than its equal-marker controls. In Smilei's upstream merging benchmark, where energy is conserved to about one part in a million, Cartesian merging triggered substantial excess while a spherical-logarithmic scheme showed smaller excess even with fewer markers (Supplementary Note~3).

When evaluated in full six-dimensional (6D) phase space, the marker floor rises steeply with diagnostic resolution (Fig.~\ref{fig:occupancy}d, inset). In the 3D pulse-driven plasma simulation of the next section, merging kept the electron count above the corresponding minimum, and every other diagnostic passed: energy balance, magnetic divergence, weight conservation, a one-dimensional (1D) energy histogram, and the exterior energy and probe checks. In the equal-marker unmerged control, initial sampling leaves most bins of the cold 6D cloud empty, creating a large initial excess that plasma heating then dilutes to the marker floor as particles spread across phase space. In the merged run, excess instead grows to about 100 times the final control as heavy merged macroparticles pack mass into the densest bins, biasing plasma heating by 2\% (Fig.~\ref{fig:occupancy}d).

\subsection*{Initial-data oracles verify nonlinear 3D plasma simulations}

Next, we tested whether initial-data oracles remain accurate and sensitive once plasma dynamics dominate the interior. We ran two 3D Smilei campaigns and evaluated the exterior references in arbitrary-precision ball arithmetic (Arb~\cite{johansson2017arb}; Tier~B; Methods and Supplementary Table~S4) so that reference rounding error is rigorously bounded.

\begin{figure}[!tbp]
\centering\includegraphics[width=\linewidth]{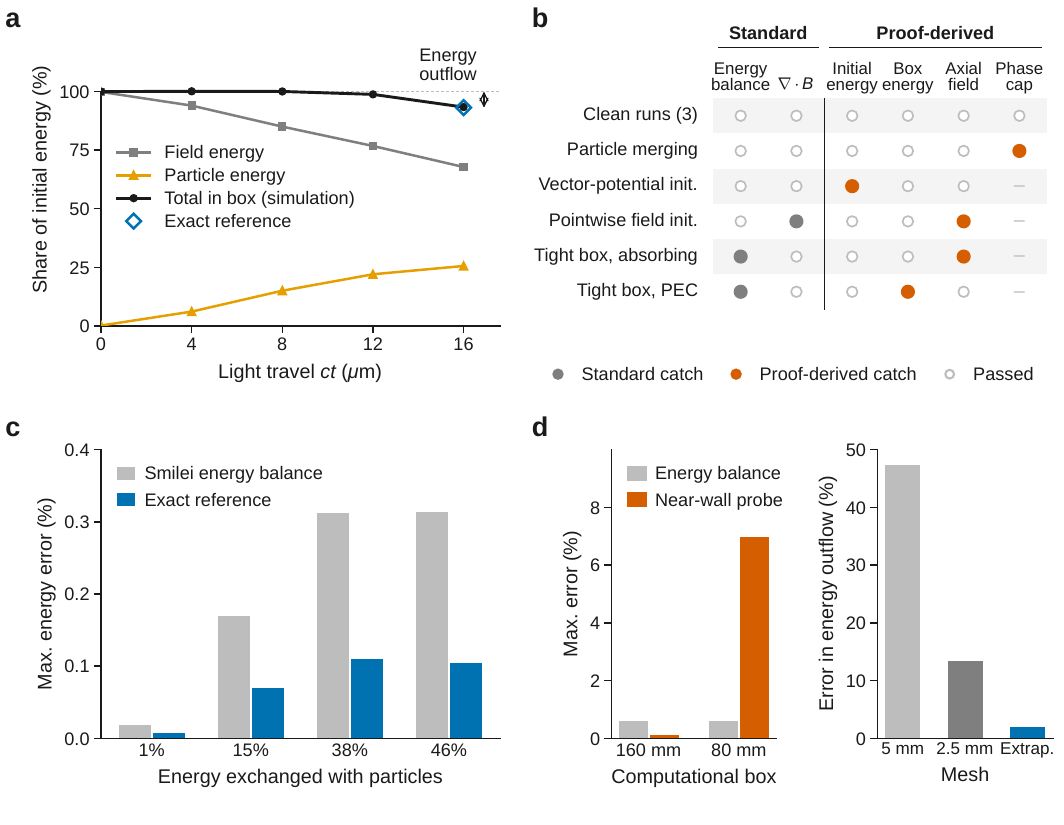}
\caption{\textbf{Exterior and phase-space diagnostics in nonlinear 3D pulse-driven and pair-plasma simulations.}
\textbf{a}, Energy budget of the 3D Smilei pulse-driven plasma run versus light travel $ct$: particles absorb roughly a quarter of the initial pulse energy while electromagnetic outflow lowers the total box energy, matching the exact exterior reference (open diamond) at $ct=16\,\mu\mathrm{m}$.
\textbf{b}, Diagnostic outcomes for three clean pulse-driven plasma runs and five single-defect perturbations across standard and proof-derived checks (filled circles, flagged; open circles, passed; dashes, not measured).
\textbf{c}, Maximum energy error in a pulse-free electron--positron plasma across four densities (indexed by the fraction of energy exchanged with particles), comparing Smilei's un-centered full-domain energy balance with the time-centered observation-box energy against the exact exterior reference.
\textbf{d}, Pulse-free pair-plasma sensitivity tests. Left: shrinking the computational box from 160~mm to 80~mm leaves the energy-balance error unchanged while raising the maximum error at a near-wall probe. Right: error in outward energy flow relative to the exact reference on 5~mm and 2.5~mm meshes and after Richardson extrapolation.}\label{fig:strong-plasma}
\end{figure}

In the first campaign, an intense toroidal electromagnetic pulse drives a compact overdense plasma. Particles absorb a quarter to a third of the initial pulse energy, and part of the remainder leaves the observation cube as outward electromagnetic flux (Fig.~\ref{fig:strong-plasma}a). Smilei's internal energy balance tracks energy over the full computational domain using solver-computed boundary fluxes. Computed from initial data alone, the exterior energy reference matched every unperturbed run to within 0.4\%.

We then introduced one numerical defect at a time into this simulation (Fig.~\ref{fig:strong-plasma}b). The two standard single-run checks, energy balance and magnetic divergence, each missed most of the defects. Both checks passed with particle merging, which the phase-space cap flagged, and with vector-potential initialization, which failed the initial-energy reference. Pointwise field sampling left energy balance intact (though magnetic divergence flagged it) and broke axial-field symmetry at exterior probes; a tight computational box failed the axial-field check for absorbing walls and the exterior energy and probe references for reflecting walls. Refining the mesh across multiple 3D runs can reveal some of these defects, yet leaves fixed-domain truncation unresolved (Fig.~\ref{fig:exterior-diag}c and Supplementary Note~2). Each diagnostic targeted a different failure mode, and together the proof-derived checks flagged every defect.

In the second campaign, counterstreaming electron--positron clouds evolved without an external pulse, interacting solely through their own fields inside the observation cube of Fig.~\ref{fig:exterior-diag}. At the highest of four densities, nearly half of the initial energy was exchanged between particles and fields (Fig.~\ref{fig:strong-plasma}c). Across all four densities, the exterior energy reference tracked the time-centered box energy without solver-computed boundary fluxes; Smilei's un-centered full-domain energy balance showed a larger residual, largely from half-step leapfrog staggering. Shrinking the computational domain left energy balance unchanged while raising the error at a near-wall probe from 0.1\% to 7\% (Fig.~\ref{fig:strong-plasma}d, left). Under mesh refinement, the simulated energy outflow converged toward the exact reference at a rate consistent with second-order accuracy (Fig.~\ref{fig:strong-plasma}d, right).

\subsection*{Formal guarantees and limits of oracle validity}

\begin{figure}[!tbp]
\centering\includegraphics[width=\linewidth]{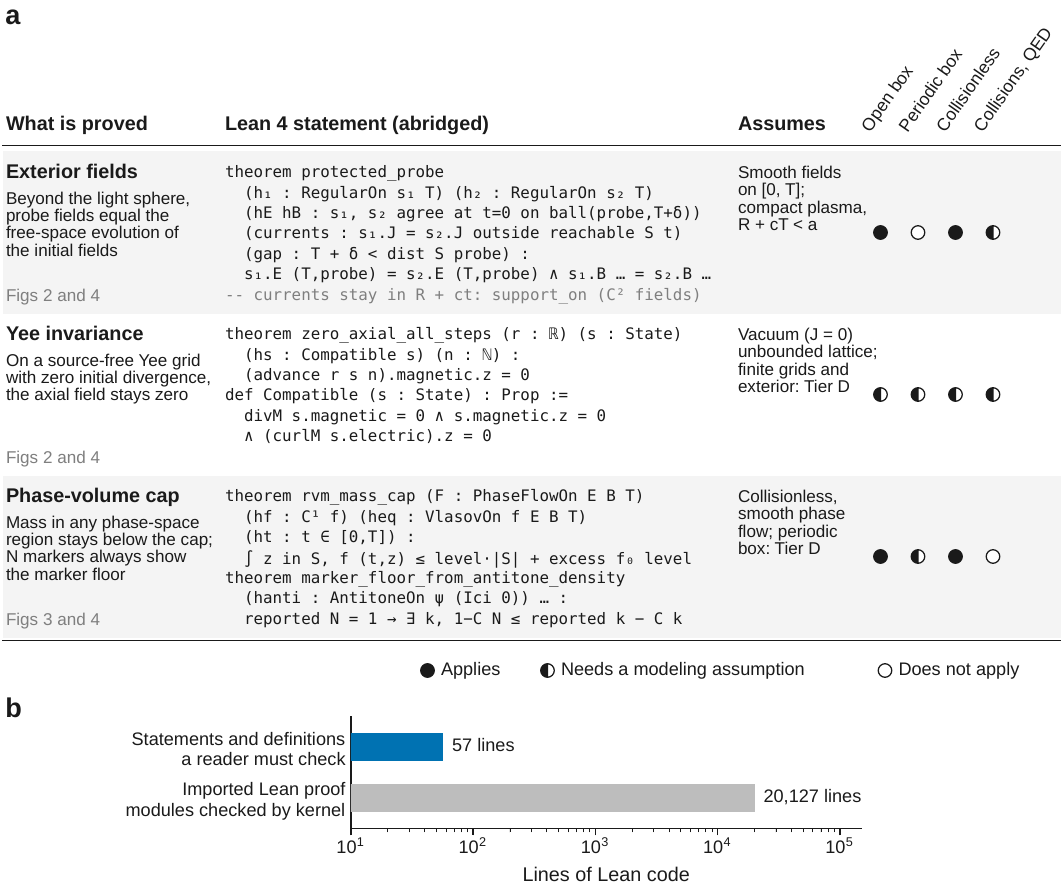}
\caption{\textbf{Scope, assumptions and Lean~4 code footprint of the formal oracles.}
\textbf{a}, Summary of the exterior-field, Yee-invariance and phase-volume oracles, listing each physical claim, abridged Lean~4 statement (full statements and definitions in Supplementary Note~1), formal assumptions, and regime applicability. Circles denote regimes where an oracle applies directly (filled), requires a Tier~D modeling assumption (half-filled), or does not apply (open).
\textbf{b}, Lean~4 code footprint on a logarithmic scale: 57 lines of statements and definitions requiring human review versus 20,127 lines of transitively imported Lean proof modules checked by the kernel across our modules and OpenAI's relativistic Vlasov--Maxwell library (excluding Mathlib; proof-term constant cones in Supplementary Note~1).}\label{fig:theorem-ledger}
\end{figure}

Because the Lean~4 kernel verifies that each bound follows from its stated premises, auditing an oracle reduces to checking that its theorem statements and supporting definitions match the intended physical model (Fig.~\ref{fig:theorem-ledger}a). At an exterior probe, two electromagnetic solutions agree whenever they share the same initial fields on the probe's backward light cone and the same currents outside the plasma's light sphere; on a source-free Yee grid, zero axial field persists when the initial discrete divergence and axial electric curl vanish. Inside the plasma, phase-space mass in any region stays below the cap of equation~\eqref{eq:occupancy}, and an $N$-marker simulation always shows at least the marker-floor excess. A fifth lemma confines plasma currents inside the expanding light sphere, which is where our assumption of twice-differentiable fields enters. These five statements and the six definitions behind them occupy 57 lines of Lean, compared with more than 20,000 lines of transitively imported proof modules beneath them, excluding Mathlib (Fig.~\ref{fig:theorem-ledger}b; full text and dependency breakdown in Supplementary Note~1). Because the Lean~4 kernel eliminates algebraic or sign errors inside the proofs, auditing those 57 lines alongside the referenced model definitions in the repository replaces checking the underlying derivations by hand.

Each oracle also carries specific applicability limits and trust boundaries (Fig.~\ref{fig:theorem-ledger}a, right). The exterior reference requires the plasma's expanding light sphere to stay inside the observation cube ($R+cT<a$) in an open or absorbing domain, so it cannot be used in periodic or plasma-filled setups. Because the probe theorem imposes conditions only on currents outside that sphere, it extends to collisional or quantum electrodynamics (QED) plasmas provided their currents remain inside the sphere. The Yee invariance theorem is proved for source-free vacuum evolution on an unbounded lattice, so applying it to finite grids or exterior regions around a plasma is a modeling assumption (Tier~D). The phase-space cap applies over arbitrary run durations in closed or outflow domains, with periodic grids treated as a Tier~D reduction, but it no longer applies when collisions, radiation reaction, ionization or QED alter phase-space density or create particles. Finally, the Arb evaluations of fluxes, probes and caps sit outside the Lean kernel (Tier~B), with the exterior flux and probe enclosures cross-checked against independent analytic expansions.

\section*{Discussion}

Formal proof libraries and AI-assisted mathematics from OpenAI~\cite{openaimath}, Google~\cite{firsching2026,cai2026cogentic}, and Anthropic~\cite{anthropic2026flt} now expand faster than human review can keep pace~\cite{xie,agmai2026,ahm2026}, confronting domain scientists with the question of how to use formal results they did not check by hand. Separating formal proof construction from physical auditing offers one workable pattern. AI coding agents can draft downstream bounds in Lean~4 while the proof kernel enforces every logical step from the stated premises, leaving physicists to inspect the theorem statements, supporting definitions, and continuum-to-discrete assumptions before implementing executable checks. Across seven automated Lean~4 campaigns run with the Simjecture workflow~\cite{simjecturepaper,simjecture}, an audit of 951 deduplicated verification receipts recorded 428 failed compilation attempts across type-checking, linter, and harness checks, and manual inspection of companion WarpX particle-pusher and Picard verification libraries confirmed 10 mathematical or rounding errors caught before publication (Supplementary Note~4 and Supplementary Table~S5). As formalized PDE and operator libraries grow, intermediate lemmas proved for major analytical theorems can be reused in the same way to test numerical codes across scientific computing.

Even within its stated domain of validity, each diagnostic leaves specific blind spots and formal gaps that guide future work. Inside the pre-causal window ($R+cT<a$), opposing kinetic and field errors within the observation cube can still cancel in the summed box energy $H(t)$. Inside the plasma, the phase-volume cap in equation~\eqref{eq:occupancy} is inherently one-sided: it flags artificial compression and merging-induced clumping but passes numerical diffusion or displaced phase-space structures. Both continuum oracles also take smooth classical solutions on $[0,T]$ as a Tier~D premise, and the Yee invariance proof does not yet include finite or periodic boundaries. In the nonlinear 3D campaigns, the Arb enclosures still run outside Lean (Tier~B); importing those outward intervals into our Lean rational-summation module would bring flux-balance assembly into the kernel, and certifying the interval evaluations themselves in Lean would close the remaining gap to Tier~A. Extending this approach to bounded discrete grids, collisional or QED invariants, and continuous-integration regression suites for production solvers remains an open next step.

\section*{Methods}

\subsection*{Formal development and evidence tiers}

All formal results are stated and proved in Lean~4.34.1~\cite{demoura2021} on top of Mathlib~\cite{mathlib2020} and OpenAI's relativistic Vlasov--Maxwell library~\cite{openaimath,openairvm}, using only Lean's standard axioms. Our downstream modules prove the exterior light-cone comparison and six-face Poynting balance, signed symmetry at mirror probes for several species, discrete Yee axial invariance, exact rational flux assembly, and the collisionless phase-space cap with its marker floor. Fourteen further theorems specialize the cap and floor to the Maxwellian, relativistic Maxwell--J\"uttner and Gaussian distributions used in plasma benchmarks and add a piecewise-envelope cap stated directly on the formal phase flow (Supplementary Notes~1 and~3). Each claim in the paper is assigned to one of four evidence tiers (Supplementary Table~S1): kernel-checked theorems (A), outward-rounded numerical enclosures computed outside Lean (B), measurements on production codes (C), and stated assumptions (D). Tier~D covers smooth classical solutions on $[0,T]$, the bridge from continuum statements to staggered grids and macroparticles, and numerical flux tables not imported into Lean.

\subsection*{Exterior references from initial data}

For each exterior campaign, we first verify in exact rational arithmetic that the light sphere of radius $R+cT$ enclosing the plasma stays strictly inside the observation cube $D$ and clear of every probe; in the three-dimensional campaigns, particle tracks also confirm that no particle leaves the box. Within this window, the Lean theorems imply that the fields on $\partial D$ and at each probe equal the free-space evolution of the full initial fields, so the probe fields and the energy flux through $\partial D$ (equation~\eqref{eq:campaign-flux}) are determined from initial data alone.

We evaluate these references with outward rounding. In the primary benchmark, the exterior reference field combines the free evolution of a toroidal magnetic pulse (evaluated with an interval-enclosed spherical-mean series) with the self-field of the charge cloud (obtained from multipole enclosures of the rigidly translated initial charge). The six-face flux is integrated over fine time bins using outward-rounded midpoint quadratures and truncated series with rigorous remainder bounds, and probe values are enclosed separately and checked against independent quadratures (Supplementary Note~2). In the nonlinear three-dimensional campaigns the same quantities are enclosed in arbitrary-precision ball arithmetic (Arb through \texttt{python-flint}~\cite{johansson2017arb}), using closed-form pulse fields or multipole expansions with rigorous tails, and each enclosure is cross-checked against an independent Taylor series or Chandrasekhar--Ferrers integral. A simulation is scored by its worst deviation from these intervals over all output times: the relative error at each probe (pointwise in the primary benchmark, normalized by the peak reference field in the pair suite) and the relative error of the box energy $H(t)$.

\subsection*{Discrete Yee projection and divergence-cleaner audit}

In source-free vacuum, the discrete Yee theorem requires zero discrete magnetic divergence, zero axial magnetic field, and zero axial discrete electric curl at initialization. In Smilei we enforce this by projecting the transverse magnetic components on each axial plane orthogonally onto the discrete divergence-free subspace in two-dimensional Fourier space; the projection preserves planar means and perturbs the initial field energy negligibly (Supplementary Note~2). To test WarpX's built-in projection divergence cleaner against the same condition, we initialized three analytic magnetic fields on a $16^3$ grid bounded by conducting walls: two with perfect-electric-conductor (PEC) boundaries (one with nonzero and one with zero wall-normal component) and one with perfect-magnetic-conductor (PMC) boundaries. Each was run with and without cleaning, in stock WarpX and in the build of Pull Request \#7586~\cite{warpxpr7586}. We report the largest normalized divergence $h\max|\nabla\cdot B|/B_{\rm ref}$ over the domain and, for PMC walls, the largest tangential field in the boundary-adjacent cell layer, with $B_{\rm ref}=m_ec/(e\cdot 1\,\mu\mathrm{m})$. For magnetic cleaning in electromagnetic solvers, the patch applies Neumann conditions to the cleaning potential at PEC walls and Dirichlet conditions at PMC walls, fills the potential's boundary ghost cells before applying the potential gradient, and leaves electrostatic solvers unchanged.

\subsection*{Phase-space cap and marker floor}

To score a distribution against equation~\eqref{eq:occupancy}, we bin particle weights, or Eulerian cell masses, on a phase-space grid of equal cell volume $v$, sort the cells by mass in descending order and compare each cumulative sum $M_k$ with the cap $C(kv)$ computed from the initial distribution. Caps come from the closed forms proved in Lean for compact-polynomial, Maxwellian, Maxwell--J\"uttner and Gaussian profiles or, for the three-dimensional Maxwell--J\"uttner distribution of Smilei's merging benchmark, from outward Arb quadrature. Cumulative sums are formed in exact rational or outward interval arithmetic; reporting the lower endpoint of the largest excess $\max_k[M_k-C(kv)]$ certifies any positive value as a cap violation. Each score is reported alongside the marker floor $1-C(Nv)$, the minimum excess enforced by $N$ point markers occupying at most $N$ cells. To separate algorithmic clumping from finite-marker noise, each merged or thinned run is compared with unmerged runs whose marker count matches the final merged count to within 3\%; for Smilei's stationary merging benchmark, the matched control is a random subsample of the unmerged initial population. Bin sizes for each campaign are listed in Supplementary Table~S6.

\subsection*{Simulation campaigns}

The particle-in-cell campaigns use pinned WarpX~\cite{warpx} and Smilei~\cite{smilei} builds, complemented by an in-house Eulerian solver. Exact code revisions and simulation parameters are listed in Supplementary Note~5 and Supplementary Table~S6. The primary exterior benchmark places a compact two-temperature positron cloud and an axisymmetric toroidal magnetic pulse inside an 80~mm half-width observation cube for $cT=48$~mm. WarpX runs use the semi-implicit electromagnetic solver~\cite{chen2020} with perfectly matched layer (PML) boundaries on 80, 120 and 160~mm computational domains; Smilei runs use the explicit Yee solver~\cite{yee} with quadratic particle shapes, Esirkepov current deposition~\cite{villasenor1992,esirkepov2001}, Silver--M\"uller boundaries and boosted anisotropic Poisson initialization~\cite{vico}, on 80 and 160~mm domains at three mesh spacings. The Eulerian beam--plasma runs evolve a relativistic beam and a counterstreaming background with a Strang-split conservative remap using the Superbee or monotonized central (MC) limiter on three successively refined meshes.

The particle-merging audits use a two-dimensional relativistic Maxwell--J\"uttner Weibel instability in Smilei, initialized by Zenitani sampling~\cite{zenitani2015} and run with gentle or aggressive Vranic merging~\cite{vranic2015}; a one-dimensional counterstreaming Weibel instability in WarpX with velocity-coincidence (VCT) or leveling thinning; and Smilei's upstream thermal-plasma merging benchmark with Cartesian and spherical merging grids. In the nonlinear three-dimensional pulse-driven suite, an intense toroidal magnetic pulse drives a localized neutral electron--proton or electron--positron plasma sampled quasi-randomly from a compactly supported six-dimensional distribution. Perturbed runs vary one numerical setting at a time: enabling particle merging, switching to vector-potential or pointwise field initialization, or shrinking the computational box to the observation cube with absorbing or reflecting walls. Six-dimensional occupancy in this suite is scored with four bins per phase-space semi-axis, on both raw and leapfrog-synchronized snapshots. The pulse-free pair suite initializes counterstreaming electron--positron clouds with only their boosted purely magnetic self-field inside the observation cube of the primary benchmark at four densities, then repeats tests on a shrunk computational domain and on coarse meshes for Richardson extrapolation.

\subsection*{Detection criteria}

Binary pass/fail outcomes in the defect-detection matrix of Fig.~\ref{fig:strong-plasma}b and the nominal exterior suites use the following thresholds. The box-energy check passes if the largest relative error, whether in Smilei's internal energy balance or against the enclosed exterior reference, stays below 0.5\%, while coarse-mesh pair runs in Fig.~\ref{fig:strong-plasma}d test convergence. Exterior probe traces, including the near-wall probe in the pair suite, pass if their relative errors stay below 3\%. The initial-energy check flags relative deviations from its reference enclosure above 0.05\%. The axial-field check flags $|B_z|$ above $10^{-6}$ in normalized units at probes where symmetry forces $B_z=0$, and the divergence check flags $h\|\nabla\cdot B\|/\|B\|$ above $10^{-8}$. Continuum Eulerian and 6D pulse-driven phase-space excess is flagged above 0.5\%, with the same 0.5\% marker-floor target; the particle-merging benchmarks in Fig.~\ref{fig:occupancy}b,c are scored directly against equal-marker controls alongside their exact marker floors, using a 1\% floor target in the Weibel audits that leveling thinning exceeds.

\subsection*{Statistics and reproducibility}

No statistical hypothesis tests were performed; all benchmarks compare solver outputs directly against the initial-data references. Every run is repeatable, using either deterministic quasi-random (Sobol) marker sampling or fixed pseudorandom seeds. In the two-dimensional Smilei and one-dimensional WarpX merging audits, gentle Vranic merging, VCT thinning, and all matched equal-marker controls were run with three independent seeds, and Fig.~\ref{fig:occupancy}b,c shows every seed value alongside the mean (ranges in Supplementary Table~S3); the unmerged baseline, aggressive-merging, leveling-thinning, and stationary Smilei benchmark runs use single seeds. Lean~4 proofs, enclosures, and benchmark analyses can be reproduced from archived inputs and pinned software revisions, and clean rebuilds from source reproduced every reported pass/fail outcome, including cross-host Smilei runs (Supplementary Table~S2). The Simjecture provenance audit (Supplementary Note~4 and Supplementary Table~S5) deduplicated 1,004 raw records by SHA-256 digest into 951 unique verification receipts and recorded 65 SHA-256 file-hash entries.

\subsection*{Use of AI tools}

AI coding agents (OpenAI Codex and Google Antigravity) were used throughout the study to draft Lean~4 proofs of downstream lemmas, write simulation inputs and analysis scripts, audit provenance logs and edit the manuscript text. Lean verification receipts for AI-drafted proofs were recorded with Simjecture~\cite{simjecturepaper,simjecture}. Every theorem was checked by the Lean kernel, and every reported number was regenerated from archived inputs by deterministic scripts. The authors reviewed all theorem statements, definitions, code, audit records and text, and take full responsibility for the content.

\section*{Data availability}
Exterior reference intervals, processed simulation outputs, provenance records, and source data for Figs.~1--5 are archived in \href{https://github.com/tomzhu0225/vlasov-maxwell-pic-benchmark/releases/tag/v2.0.2}{version 2.0.2 of the plasma solver benchmark}~\cite{plasmasolverbenchmark}. Source data for Figs.~1--5 (\texttt{Figure-1-source-data.xlsx} through \texttt{Figure-5-source-data.xlsx}) are also provided with this paper.

\section*{Code availability}
The Lean~4 proof library, arbitrary-precision and phase-space verification scripts, benchmark simulation decks, and figure-reproduction scripts are available in \href{https://github.com/tomzhu0225/vlasov-maxwell-pic-benchmark/releases/tag/v2.0.2}{version 2.0.2 of the plasma solver benchmark}~\cite{plasmasolverbenchmark}. The WarpX \texttt{ProjectionDivCleaner} fix and regression tests are available in WarpX Pull Request \#7586~\cite{warpxpr7586}.

\section*{Funding}
The authors received no external funding for this work.

\section*{Competing interests}
The authors declare no competing interests.
\bibliography{references}
\clearpage
\newgeometry{top=20mm,bottom=22mm,left=20mm,right=20mm,bindingoffset=0mm,footskip=11mm}
\pagenumbering{arabic}
\renewcommand{\thepage}{S\arabic{page}}
\setcounter{equation}{0}
\setcounter{figure}{0}
\setcounter{table}{0}
\renewcommand{\theequation}{S\arabic{equation}}
\renewcommand{\thefigure}{S\arabic{figure}}
\renewcommand{\thetable}{S\arabic{table}}
\renewcommand{\theHequation}{S\arabic{equation}}
\renewcommand{\theHfigure}{S\arabic{figure}}
\renewcommand{\theHtable}{S\arabic{table}}
\section*{Supplementary Information}
\noindent\textit{Reusing formal Vlasov--Maxwell proofs to audit plasma simulations}\\
Bowen Zhu and Jian Wu

{\small\makeatletter
\renewcommand{\@pnumwidth}{2.2em}
\renewcommand{\@tocrmarg}{3.0em}
\tableofcontents
\makeatother}
\clearpage

\section*{\phantomsection Supplementary Note 1: Evidence tiers, proof inventory and rebuild verification}
\addcontentsline{toc}{section}{Supplementary Note 1: Evidence tiers, proof inventory and rebuild verification}

Table~\ref{tab:levels-of-evidence} assigns the formal proofs, outward numerical enclosures, floating-point solver runs, and analytic assumptions used in this study to the four evidence tiers of main Fig.~1b. Across the archived verification receipts, 49 Lean~4.34.1 modules spanning 271 axiom-audited declarations, of which 16 self-contained downstream modules spanning 98 axiom-audited declarations are packaged in version~2.0.2 of the benchmark release, build on Mathlib and OpenAI's relativistic Vlasov--Maxwell library and compile with zero \texttt{sorry} under Lean's standard axioms (\texttt{propext}, \texttt{Classical.choice}, and \texttt{Quot.sound}). They cover exterior cone comparison (\texttt{LocalizedWindow}), multi-species symmetry (\texttt{SignedExterior}), discrete Yee axial invariance (\texttt{YeeAxial}), exact rational flux assembly (\texttt{FluxAssembly}, \texttt{FluxAssemblyData}), and collisionless phase-space occupancy (\texttt{PhaseOccupancy}, plus 14 production-plasma theorems in \texttt{PhaseOccupancyExt}).

\begin{table}[!htbp]
\caption{\textbf{Four-tier evidence classification matching main Fig.~1b.} The four tiers separate kernel-proved Lean~4 theorems (Tier~A), outward interval and exact rational enclosures (Tier~B), production-code measurements (Tier~C), and analytic premises (Tier~D).}
\label{tab:levels-of-evidence}
\begin{tabular}{@{}>{\raggedright\arraybackslash}p{0.23\linewidth}>{\raggedright\arraybackslash}p{0.71\linewidth}@{}}
\toprule
Evidence tier & Basis and scope \\
\midrule
Tier~A: Formal Lean~4 proof & 49 Lean~4.34.1 modules across archived receipts (271 axiom-audited declarations, zero \texttt{sorry}; 16 self-contained downstream modules spanning 98 declarations packaged in version~2.0.2) prove exterior light-cone equality, six-face Poynting balance, source-free exact-lattice Yee zero-$B_z$ invariance, signed opposite-probe symmetry, rational time-bin summation, collisionless phase-space mass caps, sharp non-increasing marker floors, and piecewise-envelope caps; continuum results take the Tier~D hypotheses as premises. \\
Tier~B: Outward interval and rational bounds & Computed outside Lean with outward rounding or exact rational arithmetic: python-flint~0.8.0 enclosures backed by FLINT~3.3.1 Arb for the nonlinear 3D pulse-driven and pulse-free pair campaigns (Supplementary Note~3), mpmath.iv and FLINT/Arb Maxwell--J\"uttner capacities, exact Python Fraction causal windows and 6D occupancy scoring, and dyadic six-face flux bins on a $2^{-87}$~J grid. \\
Tier~C: Production solver outputs & Smilei~v5.1, WarpX~26.07/26.11, pinned WarpX commit~\texttt{cf55333}, our Pull Request~\#7586 branch, and a standalone 1D1V Eulerian finite-volume solver evolve particles or phase-space distributions and fields in binary64 arithmetic. Outputs are audited with equal-marker unmerged controls, multi-seed replicates, vacuum controls, and cross-host rebuild checks. \\
Tier~D: Analytic premises and modeling bridges & Assumes a classical relativistic Vlasov--Maxwell solution with $C^2$ fields on $[0,T]$; exterior comparison reuses Lemma~10.1 of OpenAI's Vlasov--Maxwell library. Continuum-to-discrete links to Yee meshes and macroparticles, finite or periodic Yee lattice boundaries, multi-species and SI-unit liftings, multi-dimensional superlevel-set Jacobians, and periodic flow reductions are treated analytically, and Arb flux tables are not yet imported into Lean. \\
\bottomrule
\end{tabular}
\end{table}

We rebuilt the solvers, outward enclosures, and Lean proofs cleanly from source to rule out effects from cached binaries or host-specific reductions (Table~\ref{tab:rebuilds}).

\begin{table}[!htbp]
\caption{\textbf{Independent rebuilds and cross-host verification checks.} Every software layer was recompiled or regenerated from source and compared against archived benchmark outputs.}
\label{tab:rebuilds}
\begin{tabular}{@{}>{\raggedright\arraybackslash}p{0.23\linewidth}>{\raggedright\arraybackslash}p{0.28\linewidth}>{\raggedright\arraybackslash}p{0.41\linewidth}@{}}
\toprule
Rebuilt component & Build environment & Verified agreement \\
\midrule
Smilei v5.1 (local and cross-host) & Clean local build and Intel Xeon Platinum 8352V (Open~MPI~4.1.6) & Matched every input digest and all scored pass/fail checks ($\le 9.03\times 10^{-17}$~J energy, $\le 8.06\times 10^{-16}$~T probes). \\
WarpX (rebuilt archived revision) & GCC~13.4.0, binary64, FP contraction and fast-math disabled & Reproduced all 576 between-sample energy intervals, all 12 baseline sign reversals, and all 294 within-build roundtrip field arrays across the intervention controls. \\
Continuum enclosures & Single-core C/Python outward interval and python-flint~0.8.0 / FLINT~3.3.1 Arb libraries & Reproduced all derived tables and 5,551 intervals byte for byte; exterior Arb tables cross-checked against analytic series. \\
54 upstream \texttt{.olean} objects and 49 proof modules, 16 in version~2.0.2 & Lean~4.34.1 and Mathlib & Matched every object digest; kernel checked all 271 audited declarations with zero warnings and zero \texttt{sorry}. \\
\bottomrule
\end{tabular}
\end{table}

\subsection*{\phantomsection Verbatim Lean 4 theorem statements and supporting definitions}
\addcontentsline{toc}{subsection}{Verbatim Lean 4 theorem statements and supporting definitions}

Main Fig.~5a abridges the core Lean~4 theorem signatures for readability. Below we reproduce all five verbatim theorem statements together with the six supporting definitions that encode their physical hypotheses. These excerpts span 57 non-blank lines of the audited Lean sources, compared with 20,127 lines across transitively imported Lean proof modules excluding Mathlib: 2,654 lines in our 13 downstream oracle and transitive \texttt{Certified*} modules plus 17,473 lines across 41 imported modules of OpenAI's upstream Vlasov--Maxwell library, while 13 additional bundled upstream files totaling 5,099 lines are not imported. Within the kernel proof-term constant cone, the full 13-module oracle suite references 14,168 lines across 13 downstream and 25 upstream modules, and the five Fig.~5a theorems alone reference 10,539 lines across 7 downstream and 18 upstream modules. Although \texttt{RegularOn} requires only $C^1$ fields directly and \texttt{rvm\_mass\_cap} takes a $C^1$ distribution on an abstract \texttt{PhaseFlowOn}, constructing \texttt{PhaseFlowOn} via \texttt{exists\_phaseFlowOn} and proving current confinement in \texttt{CertifiedFiniteWindow.support\_on} both require $C^2$ electromagnetic fields (\texttt{ContDiffOn} $\mathbb{R}$ \texttt{2}); our Tier~D continuum premise therefore assumes $C^2$ fields on $[0,T]$.

\begingroup\leftskip=1.0em\raggedright
\noindent\textbf{1. Exterior light-cone definitions and theorems (\texttt{CertifiedFiniteWindow.lean}, \texttt{LocalizedWindow.lean}):}\\[2pt]
\texttt{structure RegularOn (s : Electromagnetic) (T : }$\mathbb{R}$\texttt{) : Prop where}\\
\hspace*{1.2em}\texttt{electric : ContDiffOn }$\mathbb{R}$\texttt{ 1 s.E (slab T)}\\
\hspace*{1.2em}\texttt{magnetic : ContDiffOn }$\mathbb{R}$\texttt{ 1 s.B (slab T)}\\
\hspace*{1.2em}\texttt{current : ContinuousOn s.J (slab T)}\\
\hspace*{1.2em}\texttt{ampere : }$\forall$\texttt{ t }$\in$\texttt{ Ioo 0 T, }$\forall$\texttt{ x : Vec, deriv (fun u => s.E (u,x)) t - curl (fun y => s.B (t,y)) x = -s.J (t,x)}\\
\hspace*{1.2em}\texttt{faraday : }$\forall$\texttt{ t }$\in$\texttt{ Ioo 0 T, }$\forall$\texttt{ x : Vec, deriv (fun u => s.B (u,x)) t + curl (fun y => s.E (t,y)) x = 0}\\[3pt]
\texttt{def reachable (S : Set Vec) (t : }$\mathbb{R}$\texttt{) (x : Vec) : Prop := }$\exists$\texttt{ y }$\in$\texttt{ S, dist y x }$\le$\texttt{ t}\\[3pt]
\texttt{theorem support\_on \{s : Solution\} \{T R : }$\mathbb{R}$\texttt{\} (pref : EnergyPrefix s T) (hT : 0 < T)}\\
\hspace*{1.2em}\texttt{(hE : ContDiffOn }$\mathbb{R}$\texttt{ 2 s.E (Icc 0 T }$\times^s$\texttt{ (univ : Set Vec)))}\\
\hspace*{1.2em}\texttt{(hB : ContDiffOn }$\mathbb{R}$\texttt{ 2 s.B (Icc 0 T }$\times^s$\texttt{ (univ : Set Vec)))}\\
\hspace*{1.2em}\texttt{(hinit : }$\forall$\texttt{ z : Phase, s.f (0,z) }$\neq$\texttt{ 0 }$\to$\texttt{ }$\lVert$\texttt{z.1}$\rVert$\texttt{ }$\le$\texttt{ R) \{t : }$\mathbb{R}$\texttt{\} (ht : t }$\in$\texttt{ Icc 0 T)}\\
\hspace*{1.2em}\texttt{(z : Phase) (hz : s.f (t,z) }$\neq$\texttt{ 0) : }$\lVert$\texttt{z.1}$\rVert$\texttt{ }$\le$\texttt{ R + t}\\[3pt]
\texttt{theorem protected\_probe \{s}$_1$\texttt{ s}$_2$\texttt{ : Electromagnetic\} \{S : Set Vec\} \{T }$\delta$\texttt{ : }$\mathbb{R}$\texttt{\} \{probe : Vec\}}\\
\hspace*{1.2em}\texttt{(h}$_1$\texttt{ : RegularOn s}$_1$\texttt{ T) (h}$_2$\texttt{ : RegularOn s}$_2$\texttt{ T) (hT : 0 }$\le$\texttt{ T) (hd : 0 < }$\delta$\texttt{)}\\
\hspace*{1.2em}\texttt{(hE : }$\forall$\texttt{ x }$\in$\texttt{ closedBall probe (T+}$\delta$\texttt{), s}$_1$\texttt{.E (0,x) = s}$_2$\texttt{.E (0,x))}\\
\hspace*{1.2em}\texttt{(hB : }$\forall$\texttt{ x }$\in$\texttt{ closedBall probe (T+}$\delta$\texttt{), s}$_1$\texttt{.B (0,x) = s}$_2$\texttt{.B (0,x))}\\
\hspace*{1.2em}\texttt{(currents : }$\forall$\texttt{ t }$\in$\texttt{ Icc 0 T, }$\forall$\texttt{ x, }$\neg$\texttt{ reachable S t x }$\to$\texttt{ s}$_1$\texttt{.J (t,x) = s}$_2$\texttt{.J (t,x))}\\
\hspace*{1.2em}\texttt{(gap : }$\forall$\texttt{ y }$\in$\texttt{ S, T+}$\delta$\texttt{ < dist y probe) :}\\
\hspace*{1.2em}\texttt{s}$_1$\texttt{.E (T,probe) = s}$_2$\texttt{.E (T,probe) }$\wedge$\texttt{ s}$_1$\texttt{.B (T,probe) = s}$_2$\texttt{.B (T,probe)}\\[6pt]
\noindent\textbf{2. Discrete Yee invariance definition and theorem (\texttt{YeeAxial.lean}):}\\[2pt]
\texttt{def Compatible (s : State) : Prop :=}\\
\hspace*{1.2em}\texttt{divM s.magnetic = 0 }$\wedge$\texttt{ s.magnetic.z = 0 }$\wedge$\texttt{ (curlM s.electric).z = 0}\\[3pt]
\texttt{theorem zero\_axial\_all\_steps (r : }$\mathbb{R}$\texttt{) (s : State) (hs : Compatible s) (n : }$\mathbb{N}$\texttt{) :}\\
\hspace*{1.2em}\texttt{(advance r s n).magnetic.z = 0}\\[6pt]
\noindent\textbf{3. Collisionless phase-volume definitions and theorems (\texttt{OAI.RVM}, \texttt{PhaseOccupancy.lean}, \texttt{PhaseOccupancyExt.lean}):}\\[2pt]
\texttt{structure PhaseFlowOn (E B : Field) (T : }$\mathbb{R}$\texttt{) where}\\
\hspace*{1.2em}\texttt{flow : Phase }$\to\mathbb{R}\to$\texttt{ Phase}\\
\hspace*{1.2em}\texttt{differential : Phase }$\to\mathbb{R}\to$\texttt{ (Phase }$\to$\texttt{L[}$\mathbb{R}$\texttt{] Phase)}\\
\hspace*{1.2em}\texttt{initial : }$\forall$\texttt{ z, flow z 0 = z}\\
\hspace*{1.2em}\texttt{ode : }$\forall$\texttt{ z t, t }$\in$\texttt{ Icc 0 T }$\to$\texttt{ HasDerivWithinAt (flow z) (lorentzField E B (t, flow z t)) (Icc 0 T) t}\\
\hspace*{1.2em}\texttt{label : }$\forall$\texttt{ z t, t }$\in$\texttt{ Icc 0 T }$\to$\texttt{ HasFDerivAt (flow }$\cdot$\texttt{ t) (differential z t) z}\\
\hspace*{1.2em}\texttt{det\_one : }$\forall$\texttt{ z t, t }$\in$\texttt{ Icc 0 T }$\to$\texttt{ (differential z t).det = 1}\\
\hspace*{1.2em}\texttt{joint : }$\forall$\texttt{ z t, t }$\in$\texttt{ Ioo 0 T }$\to$\texttt{ HasStrictFDerivAt (fun p : }$\mathbb{R}\times$\texttt{ Phase => flow p.2 p.1) ((ContinuousLinearMap.toSpanSingleton }$\mathbb{R}$\texttt{ (lorentzField E B (t, flow z t))).coprod (differential z t)) (t, z)}\\
\hspace*{1.2em}\texttt{bijective : }$\forall$\texttt{ t, t }$\in$\texttt{ Icc 0 T }$\to$\texttt{ Function.Bijective (flow }$\cdot$\texttt{ t)}\\[3pt]
\texttt{def VlasovOn (f : Particle) (E B : Field) (T : }$\mathbb{R}$\texttt{) : Prop :=}\\
\hspace*{1.2em}$\forall$\texttt{ t }$\in$\texttt{ Ioo 0 T, }$\forall$\texttt{ x v, deriv (fun r => f (r,x,v)) t + (}$\sum$\texttt{ i : Fin 3, velocity v i * spatialPartial (fun y => f (t,y,v)) x i) + (}$\sum$\texttt{ i : Fin 3, (E (t,x) + cross (velocity v) (B (t,x))) i * spatialPartial (fun w => f (t,x,w)) v i) = 0}\\[3pt]
\texttt{def excess \{Z : Type*\} [MeasurableSpace Z] (}$\mu$\texttt{ : Measure Z) (g : Z }$\to\mathbb{R}$\texttt{) (level : }$\mathbb{R}$\texttt{) : }$\mathbb{R}$\texttt{ := }$\int$\texttt{ z, max (g z - level) 0 }$\partial\mu$\\[3pt]
\texttt{theorem rvm\_mass\_cap \{E B : OAI.RVM.Field\} \{T t : }$\mathbb{R}$\texttt{\} (F : OAI.RVM.PhaseFlowOn E B T)}\\
\hspace*{1.2em}\texttt{\{f : OAI.RVM.Particle\} (hf : ContDiffOn }$\mathbb{R}$\texttt{ 1 f (Icc 0 T }$\times^s$\texttt{ (univ : Set OAI.RVM.Phase)))}\\
\hspace*{1.2em}\texttt{(heq : OAI.RVM.VlasovOn f E B T) (ht : t }$\in$\texttt{ Icc 0 T) (S : Set OAI.RVM.Phase) (level : }$\mathbb{R}$\texttt{)}\\
\hspace*{1.2em}\texttt{(hvol : volume S }$\neq\top$\texttt{) (hg : IntegrableOn (fun z => f (t,z)) S)}\\
\hspace*{1.2em}\texttt{(hex : Integrable (fun z => max (f (t,z) - level) 0)) :}\\
\hspace*{1.2em}\texttt{(}$\int$\texttt{ z in S, f (t,z)) }$\le$\texttt{ level * (volume S).toReal + excess volume (fun z => f (0,z)) level}\\[3pt]
\texttt{theorem marker\_floor\_from\_antitone\_density (}$\psi$\texttt{ : }$\mathbb{R}\to\mathbb{R}$\texttt{) (hcont : Continuous }$\psi$\texttt{)}\\
\hspace*{1.2em}\texttt{(hnonneg : }$\forall$\texttt{ t }$\ge$\texttt{ 0, 0 }$\le\psi$\texttt{ t) (hanti : AntitoneOn }$\psi$\texttt{ (Ici 0)) (v : }$\mathbb{R}$\texttt{) (hv : 0 }$\le$\texttt{ v)}\\
\hspace*{1.2em}\texttt{(N : }$\mathbb{N}$\texttt{) (hN : 0 < N) (C : }$\mathbb{N}\to\mathbb{R}$\texttt{) (hC : }$\forall$\texttt{ k, C k = }$\int$\texttt{ t in (0:}$\mathbb{R}$\texttt{)..(k:}$\mathbb{R}$\texttt{)*v, }$\psi$\texttt{ t) (hcap : C N }$\le$\texttt{ 1) :}\\
\hspace*{1.2em}\texttt{(}$\forall$\texttt{ (reported : }$\mathbb{N}\to\mathbb{R}$\texttt{), reported N = 1 }$\to\exists$\texttt{ k, 1 - C N }$\le$\texttt{ reported k - C k) }$\wedge$\\
\hspace*{1.2em}\texttt{(}$\forall$\texttt{ k : }$\mathbb{N}$\texttt{, min ((k:}$\mathbb{R}$\texttt{)/N) 1 - C k }$\le$\texttt{ 1 - C N) }$\wedge$\texttt{ min ((N:}$\mathbb{R}$\texttt{)/N) 1 - C N = 1 - C N}\par\endgroup\medskip

\section*{\phantomsection Supplementary Note 2: Exterior continuum equations, Yee invariance and WarpX divergence-cleaner audit}
\addcontentsline{toc}{section}{Supplementary Note 2: Exterior continuum equations, Yee invariance and WarpX divergence-cleaner audit}

\subsection*{\phantomsection Finite-box energy identity and primary exterior benchmark}
\addcontentsline{toc}{subsection}{Finite-box energy identity and primary exterior benchmark}

Consider a classical relativistic Vlasov--Maxwell solution with $C^2$ fields on $[0,T]$ and initial particle support inside radius $R$ enclosed by the observation cube $D=[-a,a]^3$ with $R+cT<a$. With dimensionless momentum $p$ (in units of $mc$) and Lorentz factor $\gamma(p)=\sqrt{1+|p|^2}$, the rest-subtracted box energy is
\begin{equation}
 H(t)=\int_D \frac{\epsilon_0}{2}\left(|E|^2+c^2|B|^2\right)\,\mathrm{d} x+
 mc^2\int_{\mathbb R^6}(\gamma(p)-1) f(t,x,p)\,\mathrm{d} x\,\mathrm{d} p .
 \label{eq:H}
\end{equation}
Let $(E_*,B_*)$ be any regular Maxwell field that shares the initial data $(E(0),B(0))$ and is sourced by a comparison current supported inside radius $R+ct$. Inside the backward light cone of every point on $\partial D$, as well as that of any exterior probe $p$ beyond $R+cT$, both currents vanish, so the difference field satisfies the homogeneous Maxwell energy-flux inequality and vanishes identically on $\partial D$ and at $p$. Local conservation on $[0,t]\times D$ therefore yields
\begin{equation}
 H(t)-H(0)=-\int_0^t\int_{\partial D}\frac{1}{\mu_0}(E_*\mathbin{\times}B_*)\cdot\nu\,\mathrm{d} A\,\mathrm{d} s,\qquad 0\le t\le T .
 \label{eq:flux}
\end{equation}
Our Lean~4 exterior modules formalize this source-localized comparison and six-face balance on top of Lemma~10.1 of OpenAI's Vlasov--Maxwell library, using the library's energy-flux inequality and characteristic transport.

The primary exterior benchmark uses $a=80$~mm, $cT=48$~mm ($T=160.11$~ps), and opposite probes $p_\pm=\pm(55,40,40)$~mm at $|p_\pm|=78.90$~mm. A compact two-temperature positron cloud ($N=2\times 10^{11}$) has spatial bump $b(u)=\exp[1-1/(1-u^2)]$ of half-widths $(20,15,12)$~mm inside $R<28$~mm, cold and hot momentum bumps centered at $(0.15,0.05,0.1)$ and $(2,-1.5,0.8)\,mc$ with half-widths $0.15$ and $0.3\,mc$, and $16{,}384$ Sobol markers per population in Smilei (2{,}048 per population in WarpX). The initial fields include a toroidal magnetic pulse $B_0(\hat{z}\times x/\sigma)\exp[-|x|^2/(2\sigma^2)]$ with $B_0=50$~mT and $\sigma=20$~mm. Rigid translation of the initial charge density at fixed $\beta=(0.42834779,-0.24281776,0.19095761)$ supplies the self-field comparison $(E_s,B_s)$ via $A=I-\beta\beta^\top$, $-\nabla\cdot(A\nabla\phi)=\rho_0/\epsilon_0$, $E_s=-A\nabla\phi(x-\beta ct)$, and $cB_s=\beta\times E_s$. For the free pulse in $z=x/\sigma$, $\tau=ct/\sigma$ and $u=|z|^2$, the spherical mean $\psi(\tau,z)=\tau e^{-(u+\tau^2)/2}\sum_{k\ge 0}(u\tau^2)^k/(2k+1)!$ generates exact source-free fields via $\nabla\times\nabla\times(\psi\hat{z})$ and $\nabla\times(\partial_\tau\psi\,\hat{z})$, with $B_z\equiv 0$ everywhere. Outward interval enclosures give terminal probe references $[100.0799,100.1768]\,\mu$T at $p_+$ and $[-31.8184,-31.7887]\,\mu$T at $p_-$, and nominal total outward energy transfer $[1.81645,1.81827]$~mJ.

\subsection*{\phantomsection WarpX domain truncation and field transplantation}
\addcontentsline{toc}{subsection}{WarpX domain truncation and field transplantation}

\begin{figure}[!htbp]
\centering\includegraphics[width=\linewidth]{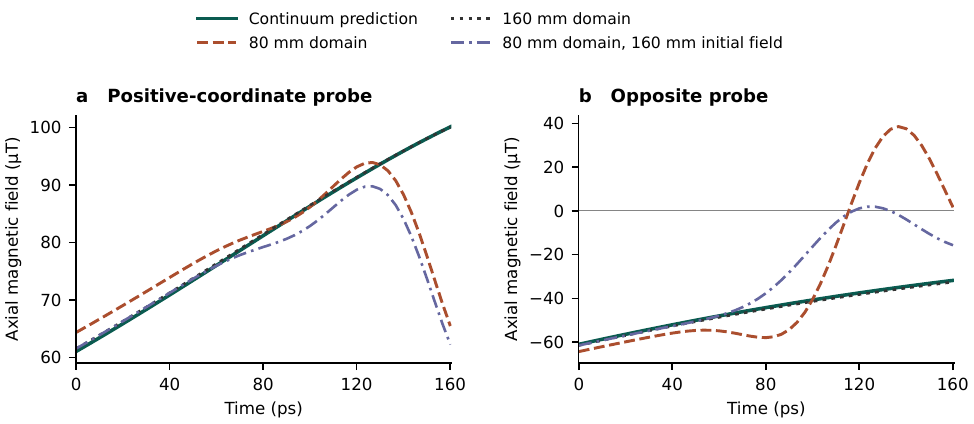}
\caption{\textbf{WarpX domain-truncation and initial-field transplantation controls at both exterior probes.} Axial magnetic field $B_z$ for the nominal 5~mm-mesh plasma run, extending main Fig.~2b to both \textbf{a}, $p_+=(55,40,40)$~mm and \textbf{b}, $p_-=(-55,-40,-40)$~mm, alongside the 80~mm-domain run initialized from 160~mm fields. Shaded bands show the continuum enclosure.}\label{fig:probes}
\end{figure}

On the 80~mm half-width WarpX domain, $p_\pm$ lie only 25~mm from the outer wall while $cT=48$~mm. All twelve baseline semi-implicit runs passed the 0.5\% box-energy check (nine also passed a 0.1\% check) yet reversed the negative $B_z$ sign at $p_-$ across 14 of 49 samples, with the nominal $N=2\times 10^{11}$ runs peaking at $+38.5\,\mu$T and ending at $+1.6\,\mu$T against the $-31.8\,\mu$T continuum reference; the backward light cone of $p_-$ reaches the 80~mm wall after 25~mm of light travel. On the 80~mm domain, tightening solver tolerances or varying perfectly matched layer thickness across nine settings (4--24 cells) left all 14 wrong signs unchanged, whereas halving cell width $h$ increased them from 14 to 15 and raised the maximum error to 253.8\%. Expanding the domain half-width $L$ from 80 to 120 and 160~mm at fixed $h=5$~mm reduced the maximum error at $p_-$ from 210.3\% (14 sign reversals) to 5.66\% and 2.53\% (zero sign reversals), giving a terminal value of $-32.5\,\mu$T on the 160~mm domain. Transplanting the 160~mm regular-grid initial fields into the 80~mm domain reduced the sign reversals from 14 to four for nominal case C07 and to five for reduced-charge and parameter controls (Fig.~\ref{fig:probes}); this control isolates initial fast-Fourier-transform truncation from dynamic wall reflection.

\subsection*{\phantomsection Discrete Yee axial invariance and transverse divergence projection}
\addcontentsline{toc}{subsection}{Discrete Yee axial invariance and transverse divergence projection}

\begin{figure}[!htbp]
\centering\includegraphics[width=\linewidth]{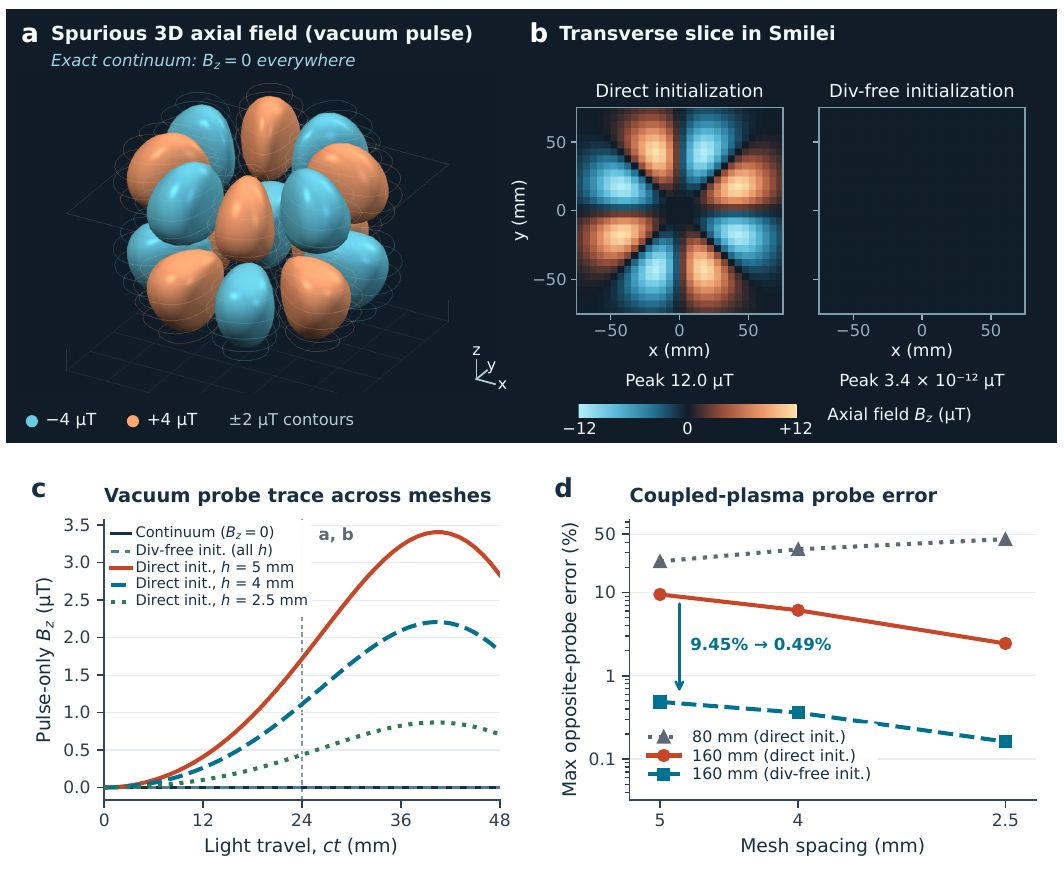}
\caption{\textbf{Spurious axial magnetic field from Yee pulse sampling in Smilei and removal by discrete projection.} \textbf{a}, Three-dimensional $\pm 4\,\mu$T isosurfaces and $\pm 2\,\mu$T contours of spurious vacuum $B_z$ at $h=5$~mm, $ct=24$~mm. \textbf{b}, Slice at $z=+20$~mm under direct sampling (peak $12.0\,\mu$T) versus 2D Fourier projection ($3.4\times 10^{-12}\,\mu$T). \textbf{c}, Vacuum probe traces across meshes. \textbf{d}, Coupled-plasma maximum relative error at $p_-$ (reproduced in main Fig.~2c).}\label{fig:smilei-lobes}
\end{figure}

Componentwise sampling of a continuum divergence-free pulse onto staggered Yee faces leaves nonzero discrete magnetic divergence $D_B B$. With $C_E$ and $C_B$ denoting the staggered Yee curls on an unbounded Cartesian lattice, commutativity of finite differences implies that the axial component of $C_E C_B B$ equals the axial difference of $D_B B$ minus the discrete Laplacian of $B_z$. Any source-free discrete vacuum state satisfying $D_B B=0$, $B_z=0$, and zero axial discrete electric curl therefore preserves $B_z=0$ identically at every leapfrog step (verified in \texttt{YeeAxial}). Subtracting $(\overline d_x,\overline d_y)(d_x\widehat B_x+d_y\widehat B_y)/(|d_x|^2+|d_y|^2)$ at nonzero wavevectors in 2D Fourier space on each axial plane projects $(\widehat B_x,\widehat B_y)$ orthogonally onto the discrete transverse divergence-free subspace while preserving planar means. In Smilei~v5.1 on the 160~mm domain, componentwise sampling left a 9.45\% coupled-plasma probe error at $p_-$ on the 5~mm mesh that was insensitive to particle count, and the pulse alone in vacuum developed sixteen alternating spurious $B_z$ lobes (Fig.~\ref{fig:smilei-lobes}). Projection drops the vacuum $z=+20$~mm slice peak from $12.0\,\mu$T to $3.4\times 10^{-12}\,\mu$T and cuts coupled-plasma 5~mm probe error from 9.45\% to 0.49\% ($|\Delta H(0)|=1.7\,\mu$J versus $\sim 350\,\mu$J for discrete $\nabla\times A$).

\subsection*{\phantomsection Upstream WarpX ProjectionDivCleaner audit and Pull Request \#7586}
\addcontentsline{toc}{subsection}{Upstream WarpX ProjectionDivCleaner audit and Pull Request \#7586}

Auditing WarpX's built-in \texttt{ProjectionDivCleaner} across WarpX~26.07, 26.11 (\texttt{93aa1ad4e}) and \texttt{development} (\texttt{c906c30bf}) exposed two coupled C++ defects in electromagnetic runs. First, \texttt{ProjectionDivCleaner::solve()} mapped \texttt{FieldBoundaryType::PEC} to \texttt{Dirichlet} ($\phi=0$) and \texttt{Neumann} (PMC) to \texttt{Neumann} ($\partial_n\phi=0$) even when cleaning $B$ directly (\texttt{m\_vector\_potential == false}), whereas $B\leftarrow B+\nabla\phi$ obeys exact magnetic duality: PEC ($B_n=0$) requires Neumann ($\partial_n\phi=0$) and PMC ($B_t=0$) requires Dirichlet ($\phi=0$). Because projection ran before \texttt{ApplyBfieldBoundary()}, the inverted PEC solve injected boundary normal flux that was immediately clamped to zero, creating a boundary-shell monopole layer. Second, \texttt{ProjectionDivCleaner.H} omitted \texttt{mlmg.setFinalFillBC(true)}, leaving boundary ghost cells of $\phi$ at zero when adding the potential gradient.

On a $16^3$ 3D PEC test whose initial field has a nonzero wall-normal component ($B_n\neq 0$; $h=1.5\,\mu$m, $B_{\rm ref}=1{,}704.51$~T; main Fig.~2d), stock WarpX (26.07 build) left full-domain $h\,\max|\nabla\cdot B|/B_{\rm ref}=0.3787$ (uncleaned $0.3952$), and on a second PEC profile with zero wall-normal field ($B_n=0$) the stock cleaner raised divergence by $+18.0\%$ in 3D ($0.5522\to 0.6516$, $16^3$; identical in 26.07 and 26.11) and $+27.5\%$ in 2D ($0.5630\to 0.7177$, $16^2$; 26.07). On a 3D PMC profile, stock WarpX (26.07 build) inflated near-wall $|B_t|/B_{\rm ref}$ from $0.1841$ to $0.5763$. Our upstream fix in WarpX Pull Request \#7586 (commit \texttt{50fbac964}) limits dual magnetic PEC/PMC mapping and pre-clean boundary clamping to electromagnetic solvers (\texttt{Yee}, \texttt{HybridPIC}) when cleaning \texttt{Bfield\_fp\_external}, enables \texttt{setFinalFillBC(true)}, rescales the solver tolerance for pre-clean clamping, and leaves external solenoid-coil handling in electrostatic \texttt{LabFrame} solvers unchanged. With this fix applied to the \texttt{development} branch in Pull Request~\#7586, 3D PEC full-domain divergence falls $6.8\times 10^{12}$-fold to $5.58\times 10^{-14}$ on the $B_n\neq 0$ profile and $7.7\times 10^{12}$-fold to $8.47\times 10^{-14}$ on the $B_n=0$ profile (retaining $\max|B|/B_{\rm ref}=0.46$ in 3D; in 2D it drops $1.2\times 10^{12}$-fold to $5.80\times 10^{-13}$). On the 3D PMC profile, first-cell near-wall $|B_t|/B_{\rm ref}$ decreases 6.2-fold to $0.0934$ ($h\,\max|\nabla\cdot B|/B_{\rm ref}=1.50\times 10^{-13}$), matching the exact continuum divergence-free value $\tfrac{1}{2}\cos^2\!\tfrac{\pi}{16}\sin\!\tfrac{\pi}{16}=0.0938$ at the first-layer cell centers $h/2$ inside the wall where $B_t=0$ identically on the wall plane, and WarpX's 2D \texttt{HybridPIC} analytical divergence checksum falls from $1{,}442.1$ to $6.45\times 10^{-11}$. With PML boundaries, where the cleaner aborts, pre-projected external openPMD fields reduce 3D axial $B_z$ 312-fold relative to 3D MLMG cleaning on periodic or conducting walls and 475-fold relative to uncleaned pointwise PML initialization, without introducing the $-0.771\%$ heating bias seen with discrete $\nabla\times A$.

\section*{\phantomsection Supplementary Note 3: Phase-space occupancy bounds, merging controls and 3D Arb campaigns}
\addcontentsline{toc}{section}{Supplementary Note 3: Phase-space occupancy bounds, merging controls and 3D Arb campaigns}

\subsection*{\phantomsection Collisionless concentration caps and sharp marker-count floor}
\addcontentsline{toc}{subsection}{Collisionless concentration caps and sharp marker-count floor}

Normalize a collisionless species distribution $f(t,z)$ on phase space $z=(x,p)$ to unit integral. For a regular Lorentz flow $F_t$ on $[0,T]$, OpenAI's upstream lemmas \texttt{PhaseFlowOn.transport} and \texttt{PhaseFlowOn.integral\_comp} give $f(t,F_tz)=f_0(z)$ and measure preservation, implying
\begin{equation}
 D_t(\lambda)=\int (f(t,z)-\lambda)_+\,\mathrm{d}z=\int (f_0(z)-\lambda)_+\,\mathrm{d}z .
 \label{eq:excess-inv}
\end{equation}
Integrating $f\le\lambda+(f-\lambda)_+$ over any measurable set $S$ of phase volume $|S|$ and minimizing over $\lambda\ge 0$ gives $\int_S f(t,z)\,\mathrm{d}z\le C(|S|):=\inf_{\lambda\ge 0}\{\lambda|S|+D_0(\lambda)\}$ (main equation~(3), proved in \texttt{PhaseOccupancy.lean}). Sorting the masses of $B$ equal-volume diagnostic cells (volume $v$ each) in descending order tests every union size $k$ in $O(B\log B)$ time; with $M_k$ denoting the cumulative mass of the $k$ heaviest cells, any positive lower endpoint of an outward enclosure of $M_k-C(kv)$ lower-bounds the mass error on that $k$-cell union. Because $N$ discrete point markers occupy at most $N$ cells of combined volume $Nv$, every $N$-marker histogram incurs excess at least $1-C(Nv)$, attained when $N$ equal-weight markers occupy $N$ distinct cells (\texttt{marker\_floor\_sharp}).

For the 1D compact beam profile $f_0\propto [1-((p-u)/w)^2]_+^3$ on periodic length $L$, $C(V)=(35s-35s^3+21s^5-5s^7)/16$ with $s=\min\{1,V/V_*\}$ and support area $V_*=2wL$. For the 6D compact cloud $f_0\propto (1-r^2)_+^3$ on semiaxes $R_i,w_i$ ($V_*=\pi^3\prod_i R_iw_i/6$, $s=\min\{1,(V/V_*)^{1/3}\}$), $C(V)=20s^3-45s^4+36s^5-10s^6$. Rational Machin-$\pi$ brackets then give exact 0.5\%-tolerance marker floors $N_{\min}=208$, $13{,}296$, and $151{,}441$ for $h_{\rm bin}=2,4,6$ diagnostic bins per semi-axis. Fourteen additional theorems in \texttt{PhaseOccupancyExt.lean} (Tier~A) prove a general marker floor for non-increasing (antitone) profiles, a piecewise-envelope cap directly on \texttt{OAI.RVM.PhaseFlowOn}, and closed-form caps for 2D Maxwellian, 2D relativistic Maxwell--J\"uttner, and 6D Gaussian distributions. For Smilei's upstream merging deck, 3D Maxwell--J\"uttner capacities are evaluated with outward-rounded FLINT~3.3.1 Arb ball arithmetic (Tier~B). Table~\ref{tab:merging-summary} summarizes all continuum remapping and production particle-merging audits plotted in main Fig.~3.

\begin{table}[!htbp]
\caption{\textbf{Phase-space remapping and particle-merging audits versus equal-marker controls (main Fig.~3).} Excess reports the outward-enclosure lower endpoint of $\max_k[M_k-C(kv)]$; equal-$N$ controls are unmerged runs whose marker counts match the final merged counts to within 3\% (for Smilei's stationary upstream deck, a random subsample of the unmerged initial population); MC, monotonized central limiter; VCT, velocity-coincidence thinning; sph-log, spherical-logarithmic; $U_{\rm elm}$, electromagnetic energy; raw/sync, difference between raw and leapfrog-synchronized particle outputs.}
\label{tab:merging-summary}
\begin{tabular}{@{}>{\raggedright\arraybackslash}p{0.23\linewidth}>{\raggedright\arraybackslash}p{0.18\linewidth}>{\raggedright\arraybackslash}p{0.15\linewidth}>{\raggedright\arraybackslash}p{0.15\linewidth}>{\raggedright\arraybackslash}p{0.14\linewidth}@{}}
\toprule
Benchmark and code & Setting / markers & Excess (\%) & Equal-$N$ / floor & Global check \\
\midrule
1D1V Eulerian ($\omega_{\rm p}T=36$) & Superbee ($64\times 160$, $128\times 320$, $256\times 640$) & 1.338, 1.526, 1.064 & MC: 0.000--0.008 & $\Delta E=0.035$--$0.065\%$ ($L^1=0.069$) \\
Smilei 2D2V Weibel ($\omega_{\rm p}T=23.9$) & Vranic gentle, $12{,}624$, 3 seeds & 5.55, range 5.29--5.89 & 0.96, range 0.95--0.98 / 0.00 & Energy bal.\ $1.05$--$1.08\%$ \\
Smilei 2D2V Weibel ($\omega_{\rm p}T=23.9$) & Vranic agg., $3{,}075$, 1 seed & 26.03 & 6.64, range 6.53--6.72 / 0.92 & $B_z$ err 35.2\%, $U_{\rm elm}$ 7.6\% \\
WarpX 1D2V Weibel ($\omega_{\rm p}T=30$) & VCT, $4{,}421$, 3 seeds & 5.73, range 5.22--6.01 & 2.78, range 2.34--3.11 / 0.09 & $\Delta E=0.18$--$0.19\%$; leveling 1.1\%, 1 seed, floor 2.79\% \\
Smilei upstream merging deck & Cartesian, $83{,}562$, 1 run & 35.08; sph-log 22.87 & 8.68 / 0.75; sph-log 9.47 / 1.04 & Energy bal.\ $1.14\times 10^{-4}\%$ \\
Smilei 3D pulse-driven (6D) & Vranic ($16{,}304$, $h_{\rm bin}=4$) & 7.146 (raw/sync $\Delta 0.040$) & 0.070 / 0.063 & Energy bal.\ $0.308\%$, $+1.99\%$ heat \\
\bottomrule
\end{tabular}
\end{table}

In the continuum Eulerian runs, Superbee kept $\max f/\max f_0\le 0.9998$, drifted seven times less in energy than MC on the $64\times 160$ mesh, and also had slightly lower global $L^1$ error than MC on the $128\times 320$ mesh. Its non-monotone excess across the three meshes (1.34\%, 1.53\%, 1.06\%) is consistent with gradient steepening across newly resolved filaments. For the Smilei 2D Weibel runs, gentle merging exceeded its three-seed equal-marker control by a factor of 5.8, whereas aggressive merging exceeded its own control by a factor of 3.9 (28 times its marker floor), raised the relative $L^2$ error in $B_z$ against a 128-ppc reference from 30.8\% in the equal-marker control to 35.2\%, and more than doubled the peak electromagnetic-energy error from 3.55\% to 7.60\%, compared with 0.46\% in the unmerged 32-ppc baseline. WarpX VCT yielded twice the excess of its equal-marker control; its 3D random azimuthal rotation introduces out-of-plane momentum in 1D2V runs, yet macroscopic field errors remain comparable to those of unmerged controls. On Smilei's upstream merging benchmark, Cartesian merging gave four times the equal-marker excess, whereas spherical-logarithmic merging was 12 percentage points lower. Because that setup is stationary, a 1D energy histogram compared against the equilibrium distribution also reveals the resulting comb artifacts, whereas the phase-volume cap applies unchanged to evolving plasmas. Raising the 6D pulse-driven bin count per semi-axis from $h_0$ to $h_{\rm bin}$ multiplies the cell count by $(h_{\rm bin}/h_0)^6$ and accounts for the 64-fold ($h_{\rm bin}=2\to 4$) and $1.5^6\approx 11.4$-fold ($4\to 6$) floor increases. Merging reduced the electron marker count from 131,072 to 16,304, still above the 13,296-marker floor at $h_{\rm bin}=4$. In the equal-marker control, initial Sobol sampling placed all 16,384 markers into 7,326 bins of the cold 6D cloud, giving 6.88\% excess at full support, and plasma heating then dispersed them across 16,185 bins, lowering the excess to 0.07\%, near its 0.06\% floor, whereas in the merged run heavy merged macroparticles packed 39.6\% of the mass into the 801 densest bins, raising the excess from 0.08\% to 7.15\% (102 times the control and 113 times its floor). Throughout that merged run, built-in energy balance, exterior energy and probe checks, discrete magnetic divergence, weight conservation and the 1D relativistic energy histogram (total variation 0.049) all passed.

\subsection*{\phantomsection Nonlinear 3D Arb campaigns}
\addcontentsline{toc}{subsection}{Nonlinear 3D Arb campaigns}

Two strongly nonlinear 3D Smilei~v5.1 campaigns were validated against outward-rounded arbitrary-precision ball enclosures computed directly in \texttt{python-flint}~0.8.0 (FLINT~3.3.1 Arb, Tier~B) and cross-checked against independent Taylor expansions and Chandrasekhar--Ferrers integrals (Table~\ref{tab:arb-campaigns}; main Fig.~4).

\begin{table}[!htbp]
\caption{\textbf{Nonlinear 3D pulse-driven and pulse-free pair campaigns validated by FLINT~3.3.1 Arb (main Fig.~4).} Exact rational arithmetic verifies strict causal separation ($R+cT<a$), and particle tracking confirms zero particle escape. Abbreviations: abs., pulse energy absorption; $H_{\rm cont}$, relative box-energy residual against the Arb continuum reference; $H_0$, initial box energy; $\Delta H(48)$, electromagnetic energy outflow over $cT=48$~mm; eq-$N$, equal-marker unmerged control; 1D TV, 1D energy-histogram total variation; SM, Silver--M\"uller; PEC, perfect electric conductor.}
\label{tab:arb-campaigns}
\begin{tabular}{@{}>{\raggedright\arraybackslash}p{0.25\linewidth}>{\raggedright\arraybackslash}p{0.21\linewidth}>{\raggedright\arraybackslash}p{0.20\linewidth}>{\raggedright\arraybackslash}p{0.21\linewidth}@{}}
\toprule
Campaign and run & Physical regime & Proof-derived check & Standard diagnostic \\
\midrule
Pulse-driven $e^-p^+$ clean ($h=0.75, 1.0\,\mu$m) & $B_0/B_r=1.5$, $\omega_{\rm p}T=25.31$, abs.\ 25.5--25.8\% & Arb $H_{\rm cont}$ 0.215--0.349\%, $B_y$ 0.67--1.40\% & Energy bal.\ 0.25--0.31\% (6.5--6.6\% outflow) \\
Pulse-driven $e^-e^+$ clean / vacuum & $\omega_{\rm p}T=22.63$, abs.\ 34.47\% & Arb $H_{\rm cont}$ 0.166\% / 0.186\% & Energy bal.\ 0.41\% / 0.09\% (6.7\% outflow) \\
Pulse-driven: merging perturbation & $131{,}072\to 16{,}304$ ($+1.99\%$ heating bias) & 6D excess 7.146\% (102$\times$ eq-$N$) & Energy bal.\ 0.308\%, $\nabla_h\cdot B=0$, 1D TV 0.049 \\
Pulse-driven: discrete $\nabla\times A$ & Discrete-curl shift $-0.346\%$ ($\mathcal{O}(h^2/\sigma^2)$) & Arb $H_0$ error 0.346\% & Energy bal.\ and $\nabla_h\cdot B$ pass \\
Pulse-driven: pointwise $B$ & $L=40\,\mu$m SM & Norm.\ Yee $|B_z|$ $5.1\times 10^{-5}$ & Energy bal.\ 0.31\%, $\nabla_h\cdot B=1.67\times 10^{-4}$ \\
Pulse-driven: tight walls & $L=24\,\mu$m SM vs PEC & SM norm.\ Yee $|B_z|$ $9.2\times 10^{-4}$; PEC $H_{\rm cont}$ 3.61\%, $B_y$ 36.0\% & Energy bal.: SM 1.02\%, PEC 2.11\% \\
Pulse-free $e^-e^+$ ($\beta_x=0.70$, $R=20$~mm) & $\omega_{{\rm p,pair}}T=2.81$, $8.87$, $14.03$, $19.84$; 1.5--46.0\% kinetic & Arb $H_{\rm cont}$ 0.008--0.110\% across all four runs & Energy bal.\ 0.02--0.31\% (0.06--1.26\% outflow) \\
Pulse-free: tight $L=80$~mm & $L=80$ vs 160~mm ($h=4$~mm) & Near-wall probe jump 66.9$\times$ (0.104\% to 6.973\%) & Energy bal.\ 0.606\% vs 0.603\% \\
Pulse-free: coarse meshes & 2nd-order Richardson & $\Delta H(48)$ within 0.017--0.018\% of $H_0$ & Energy bal.\ 0.60--2.19\% \\
\bottomrule
\end{tabular}
\end{table}

In the pulse-driven campaign, the initial box energy matched its Arb enclosure to within $3\times 10^{-7}$ relative error and exterior probes agreed within 1.4\%. Clean-run Arb box-energy residuals (0.17--0.35\%) were comparable to or smaller than Smilei's internal energy-balance error (0.25--0.41\%). Pointwise magnetic sampling introduced a nonzero initial discrete divergence and increased the spurious axial field $|B_z|$ at exterior probes by a factor of $4.8\times 10^5$ relative to the clean run. Under discrete $\nabla\times A$ initialization, the initial energy deficit matched the exact second-order finite-difference prediction. In the pulse-free pair campaign, raising density increased the peak-to-peak kinetic energy exchanged between particles and fields from 1.5\% to 46\% of the initial total energy $H_0$; at the highest density, newly generated electric-field energy peaked at 34\% of $H_0$ and magnetic energy fell by 84\% of its initial value (51\% of $H_0$). Across all four densities, the time-centered box-energy residuals against the Arb reference (0.01--0.11\%) were smaller than Smilei's un-centered full-domain energy-balance residual (0.02--0.31\%), of which 72--88\% is attributable to the half-step leapfrog kinetic-energy staggering term $\tfrac{1}{2}\Delta t|\mathrm{d}K/\mathrm{d}t|/H_0$ spanning 0.014--0.276\%. Halving the computational domain half-width from 160 to 80~mm shifted the internal energy-balance error only from 0.60\% to 0.61\% while raising the near-wall probe error 66.9-fold. Second-order Richardson extrapolation combining the 5 and 2.5~mm meshes agreed with the Arb energy outflow $\Delta H$ to within 1.99\% and 1.56\% at the two highest densities (0.017--0.018\% of $H_0$). Unlike discretization error, fixed-domain truncation was not removed by mesh refinement alone: in the 80~mm exterior tests (Supplementary Note~2), halving $h$ in WarpX increased the sign reversals from 14 to 15, and refining the mesh in Smilei likewise left the truncation error unresolved (main Fig.~2c).

\section*{\phantomsection Supplementary Note 4: Provenance audit of Simjecture Lean~4 campaigns}
\addcontentsline{toc}{section}{Supplementary Note 4: Provenance audit of Simjecture Lean~4 campaigns}

We audited Lean~4 kernel outcomes archived in \texttt{provenance\_summary.json} across all seven Simjecture campaign roots, which cover companion WarpX Boris-pusher, Picard-gate, Lipschitz-pulse, Euler, and optimality proof libraries. Removing 12 copied staging records and byte-identical duplicates identified by SHA-256 reduced 1,004 raw records to 951 unique verification receipts totaling 6.58~h of kernel wall time: 498 passing proofs, 25 intentional mutation controls that tested harness rejection, and 428 organic compilation rejections. Median wall time was 9.14~s (interquartile range 4.53--21.42~s) for organic rejections versus 14.67~s (8.29--29.15~s) for passes.

Of the 428 organic rejections, 150 (35.0\%) were style-linter failures under \texttt{warningAsError}, 52 (12.1\%) were environment, syntax, timeout, audit, or other harness errors, and 226 (52.8\%) were mathematical or type failures, most of them routine tactic retries. We manually inspected the archived failed drafts, compiler logs, and passing repairs from the companion WarpX Boris-pusher, Lipschitz-pulse, and Picard-gate campaigns and pinned 65 SHA-256 hash entries spanning 61 unique source, log, and receipt files in \texttt{result4\_catches.json}; 49 entries cover Table~\ref{tab:all-catches} and 12 cover two companion checks. Manual audit confirmed ten substantive errors (Table~\ref{tab:all-catches}): six macro-scale mathematical errors (entries 1--6) and four numerical-enclosure or rounding errors (entries 7--10). We excluded all 25 intentional mutation controls along with three candidates that proved to be false leads or non-plasma drafts. These ten catches come from the companion WarpX libraries, separate from the three main-text oracle families of main Fig.~5; they show the errors intercepted by the Lean~4 kernel under the same AI-assisted workflow.

\begin{table}[!htbp]
\caption{\textbf{Ten audited AI-drafted errors intercepted by Lean~4 in companion WarpX formalization campaigns.} Entries 1--6 are macro-scale mathematical errors; entries 7--10 are numerical-enclosure and rounding errors. All 65 SHA-256 hash entries across 61 unique source, compiler-log, and receipt files are pinned in the accompanying provenance archive.}
\label{tab:all-catches}
\begin{tabular}{@{}>{\raggedright\arraybackslash}p{0.04\linewidth}>{\raggedright\arraybackslash}p{0.10\linewidth}>{\raggedright\arraybackslash}p{0.38\linewidth}>{\raggedright\arraybackslash}p{0.34\linewidth}@{}}
\toprule
\# & Archive ID & Rejected AI-drafted error & Impact if uncaught \\
\midrule
1 & A4 & Inserted magnetic pulse Lipschitz bound into electric Boris shift slot & Bound $2\times$ too small on all 4 grids; omits grid electric term \\
2 & A2 & Unannotated rational momentum caps parsed as natural-number division & Shrinks momentum-ball norm 24.4\% and 3D box volume 65.5\% \\
3 & A3 & Interval-squaring lemma required $L\le 0$, failing on 12 positive $|u|^2$ bounds & Collapses $\gamma\ge 1.0503\to 1.0$; inflates Boris $1/\gamma$ bound $+5.03\%$ \\
4 & A1 & Applied nonnegative-numerator division rule to negative interval $[-32,0]/64$ & Reverses lower bound, up to $+1.59\%$ on an illustrative $[63,64]$ test (resolved by sign-safe reordering) \\
5 & A5 & Fabricated five nonexistent upstream Vlasov--Maxwell library identifiers & Disconnects discrete gate from PDE (blocked by \texttt{autoImplicit false}) \\
6 & A6 & Weakened Gr\"onwall rate premise to $0\le C$ on branch dividing by $C$ & Drops forcing growth at $C=0$ under Lean's $0/0=0$ convention (fixed by requiring $0<C$) \\
7 & A7 & Omitted the $+1$ endpoint term in a $1{,}000$-substep rational Picard budget & Unsound rational budget overshoot of $+6.0$~nJ \\
8 & A8 & Substituted binary64 0.01 for exact rational $1/100$ in spatial ceiling & Violates exact upper bound by $+2.08\times 10^{-19}$~m \\
9 & A9 & Substituted 193-bit outward dyadic ceiling for exact timestep bound & Unsound upper-endpoint drift at dyadic precision \\
10 & A10 & Substituted 80-bit outward dyadic ceiling for timestep bound (pre-receipt draft) & Unsound timestep ceiling (caught before finalizing receipt) \\
\bottomrule
\end{tabular}
\end{table}

\section*{\phantomsection Supplementary Note 5: Simulation and diagnostic parameters}
\addcontentsline{toc}{section}{Supplementary Note 5: Simulation and diagnostic parameters}

Table~\ref{tab:sim-params} lists the parameters of every campaign in the main text. Lengths and times follow the units of each input deck: millimeters for the exterior and pair campaigns, micrometers for the pulse-driven suite (normalized length unit $1\,\mu$m) and divergence-cleaner tests, and plasma units (electron skin depth $c/\omega_{\rm p}$, electron Debye length $\lambda_{\rm De}$, and inverse plasma frequency $\omega_{\rm p}^{-1}$) elsewhere. Marker counts are per species or per population, as stated.

\begin{table}[!htbp]
\caption{\textbf{Simulation and diagnostic parameters (main Figs.~2--4 and Methods).} PML, perfectly matched layer; FFT, fast Fourier transform; 1D1V, one spatial and one momentum dimension; MC, monotonized central limiter; VCT, velocity-coincidence thinning; PEC, perfect electric conductor; PMC, perfect magnetic conductor.}
\label{tab:sim-params}
\ifdim\textheight>620pt\footnotesize\else\fontsize{6.4}{7.2}\selectfont\fi
\setlength{\tabcolsep}{3.5pt}\renewcommand{\arraystretch}{0.98}\begin{tabular}{@{}>{\raggedright\arraybackslash}p{0.13\linewidth}>{\raggedright\arraybackslash}p{0.22\linewidth}>{\raggedright\arraybackslash}p{0.25\linewidth}>{\raggedright\arraybackslash}p{0.33\linewidth}@{}}
\toprule
Campaign & Code, solver and boundaries & Domain, mesh and time & Particles, physics and diagnostics \\
\midrule
Exterior benchmark, Smilei (Fig.~2a,c) & Smilei~v5.1; explicit Yee; quadratic shape; Esirkepov deposition; Silver--M\"uller field boundaries; particles removed at walls & Half-width 80 and 160~mm (120~mm control); $h=5$, 4, 2.5~mm; $c\Delta t=1$~mm (0.5~mm control); $cT=48$~mm, 49 outputs & 16{,}384 Sobol markers per population (cold and hot); positron cloud $N=2\times10^{11}$ with cuboid half-widths $(20,15,12)$~mm; pulse $B_0=50$~mT, $\sigma=20$~mm; probes $\pm(55,40,40)$~mm \\
Exterior benchmark, WarpX (Fig.~2a,b) & WarpX (\texttt{cf55333}); $\theta$-implicit electromagnetic ($\theta=0.5$), Newton--GMRES; linear shape; PML 8 cells (4--24 in controls); absorbing particle walls & Half-width 80, 120, 160~mm; $h=5$~mm (2.5~mm control); $c\Delta t=1$~mm; 48 steps & 2{,}048 markers per population; FFT self-field initialization; same physical setup and probes as Smilei \\
Eulerian beam--plasma (Fig.~3a) & In-house 1D1V conservative remap, Strang splitting; Superbee or MC limiter; periodic & $x\in[0,6)\,c/\omega_{\rm p}$, $p\in[-2,4.5]\,mc$; meshes $64\times160$, $128\times320$, $256\times640$ with $\omega_{\rm p}\Delta t=1/16$, 1/32, 1/64; $\omega_{\rm p}T=36$ & Beam density 0.2, $p_b=2\,mc$, half-width $0.35\,mc$; background 0.8; cap scored on native cells \\
Weibel merging, Smilei (Fig.~3b) & Smilei~v5.1, 2D; quadratic shape; periodic & $6\times6\,c/\omega_{\rm p}$, $32\times32$ cells; $\omega_{\rm p}\Delta t\approx 0.060$; 400 steps ($\omega_{\rm p}T=23.9$) & 32{,}768 markers per beam; Maxwell--J\"uttner inverse temperature $\alpha=8$, drift $\gamma_b=2$; Vranic merging, gentle every 40 steps on $4\times4\times1$ momentum cells, three seeds, or aggressive every 20 steps on $2\times2\times1$, one seed, with three-seed equal-$N$ controls and a single-seed 32k baseline; cap bins: $6\times6$ in space, widths $0.25\times0.5\,mc$ in $(p_x,p_y)$; floor tolerance 1\% \\
Weibel thinning, WarpX (Fig.~3c) & WarpX~26.11, 1D ($z$); quadratic shape; periodic & $L=4\,c/\omega_{\rm p}$, 64 cells; $\omega_{\rm p}\Delta t\approx 0.031$; 960 steps ($\omega_{\rm p}T=30$) & 32{,}768 markers per beam; dimensionless proper velocity $u_x=\pm1.5$ ($u=p/mc$), thermal spread 0.10 ($x$) and 0.05 ($z$); VCT with three seeds and three-seed equal-$N$ controls, or leveling thinning with one seed; cap bins: 16 in $z$, widths $0.04\times0.02$ in $(u_x,u_z)$; floor tolerance 1\% \\
Smilei merging benchmark (Fig.~3c) & Smilei~v5.1 upstream deck \texttt{tst1d\_17}; Cartesian and spherical (linear, logarithmic) merging grids & 256 cells, $\Delta x=\lambda_{\rm De}/16$; $\Delta t=0.95\,\Delta x/c$; 50 steps & 20{,}480 markers per cell (5.24 million per species); $T_e=100$~keV; cap bins: 32 in $x$, width $0.25\,mc$ in each momentum component; 3D Maxwell--J\"uttner cap by Arb \\
3D pulse-driven plasma (Fig.~4a,b; Fig.~3d) & Smilei~v5.1, 3D; quadratic shape; Silver--M\"uller (reflecting for the PEC variant) & Observation cube $[-24,24]^3\,\mu$m; computational half-width $40\,\mu$m ($42\,\mu$m at $h=0.75\,\mu$m; $24\,\mu$m for tight boxes); $h=1\,\mu$m, $c\Delta t=0.25\,\mu$m ($0.75$ and $0.20\,\mu$m for the fine run); $cT=16\,\mu$m, outputs every $4\,\mu$m & 131{,}072 markers per species (32{,}768 in non-merging perturbed runs and $e^-e^+$); compact 6D distribution, $R=6\,\mu$m, momentum semi-axis $0.20\,m_ec$; $e^-p^+$ or $e^-e^+$; co-located toroidal magnetic pulse $B_0/B_r=1.5$, peak $eA_z/m_ec=9.0$, $\sigma=6\,\mu$m; Vranic merging (merging variant: every 8 steps, $4^3$ momentum cells, \texttt{merge\_\allowbreak accumulation\_\allowbreak correction=False}); 6D bins: four per semi-axis \\
Pulse-free pair plasma (Fig.~4c,d) & Smilei~v5.1, 3D; quadratic shape; Silver--M\"uller & Half-width 160~mm (80~mm truncation run); $h=2.5$~mm, $c\Delta t=0.25$~mm (coarse: $h=5$~mm with $c\Delta t=1$~mm, and $h=4$~mm with $c\Delta t=0.5$~mm); $cT=48$~mm, outputs every 4~mm & 32{,}768 markers per species; cloud semi-axes $(20,16,16)$~mm, drift $\pm1\,mc$ (mean $\beta_x=0.70$, momentum half-width $0.4\,mc$); $\omega_{\rm p,pair}T=2.8$, 8.9, 14.0, 19.8; purely magnetic boosted self-field only; probes $\pm(55,50,20)$~mm, near-wall $\pm(70,50,20)$~mm \\
Divergence cleaner (Fig.~2d) & WarpX~26.07 and 26.11 (stock) and Pull Request~\#7586 build (\texttt{50fbac964}) & $16^3$ cells on $[-12,12]^3\,\mu$m ($16^2$ on $[-12,12]^2\,\mu$m in 2D), $h=1.5\,\mu$m; four steps & Analytic fields: PEC with $B_n\neq0$, PEC with $B_n=0$, PMC; $B_{\rm ref}=1{,}704.51$~T \\
\bottomrule
\end{tabular}
\end{table}
\end{document}